\documentclass[aps,
               pra,
               showpacs,
               amssymb,
               nofootinbib,
               superscriptaddress,
               twocolumn,
               longbibliography]{revtex4-2}
\usepackage[ansinew]{inputenc}
\usepackage{bbold}
\usepackage{bbm}
\usepackage{bm}
\usepackage{amsbsy}
\usepackage{amsthm}
\usepackage{amssymb}
\usepackage{amsfonts}
\usepackage{amsmath, mathtools}
\usepackage{dsfont}
\usepackage{graphicx}
\usepackage{epsfig}
\usepackage{epstopdf}
\usepackage{dsfont}
\usepackage{mathrsfs} 
\usepackage{multibib}
\usepackage[dvipsnames]{xcolor}
\usepackage[colorlinks]{hyperref}
\makeatletter
\newcommand\org@hypertarget{}
\let\org@hypertarget\hypertarget
\renewcommand\hypertarget[2]{%
  \Hy@raisedlink{\org@hypertarget{#1}{}}#2%
  }
\makeatother
\usepackage[figure,table]{hypcap}
\usepackage{MnSymbol}
\usepackage{enumerate}
\usepackage{float}
\usepackage{comment}
\usepackage{braket}
\usepackage{subcaption}
\usepackage{ragged2e}
\usepackage{orcidlink}

\hypersetup{
	bookmarksnumbered,
	pdfstartview={FitH},
	citecolor={darkgreen},
	linkcolor={darkred},
	urlcolor={darkblue},
	pdfpagemode={UseOutlines}}
\definecolor{darkgreen}{RGB}{50,190,50}
\definecolor{darkblue}{RGB}{0,0,190}
\definecolor{darkred}{RGB}{238,0,0}
\usepackage{soul}

\newcommand{\ketbra}[2]{\ensuremath{|{#1}\rangle\!\langle{#2}|}}

\newcommand{\tr}{\textnormal{Tr}}
\newcommand{\djj}{d\kern-0.4em\char"16\kern-0.1em}

\makeatletter
\renewcommand{\p@subsection}{}
\renewcommand{\p@subsubsection}{}
\makeatother

\begin{document}

\title{
Compressibility of genuine multipartite entanglement under the Hadamard map
}
\author{Kl{\'a}ra Baksov{\'a}\,\orcidlink{0009-0009-8944-6044}}
\email{klara.baksova@matfyz.cuni.cz}
\affiliation{Faculty of Mathematics and Physics, Charles University, Ke Karlovu 3, 121 16 Praha 2, Czech Republic}
\affiliation{Technische Universit{\"a}t Wien, Atominstitut \& Vienna Center for Quantum Science and Technology (VCQ),  Stadionallee 2, 1020 Vienna, Austria}

\author{Lisa T. Weinbrenner}
\affiliation{Naturwissenschaftlich-Technische Fakult\"at, Universit\"at Siegen, Walter-Flex-Stra{\ss}e 3, 57068 Siegen, Germany}

\begin{abstract}
    One of the most counterintuitive effects in the examination of quantum states is the phenomenon of superactivation, which describes the fact that a quantum state, which is useless for a specific task, may become useful when one considers multiple copies of it.
    This effect can be observed in the case of genuine multipartite entanglement, where local projections from multiple copies to the single-copy Hilbert space could significantly simplify the practical accessibility of its superactivation. We investigate how such projections behave in the limit of many copies, using the Hadamard map as an example, and studying different state families. We show that, in this scheme, for a fixed map, an optimal number of copies exists, beyond which the obtained entanglement decreases, highlighting the differences between entanglement distillation and local projection schemes.
\end{abstract}

\maketitle

\section{Introduction}

    \begin{figure}[t]
        \centering
        \includegraphics[width=1\linewidth]{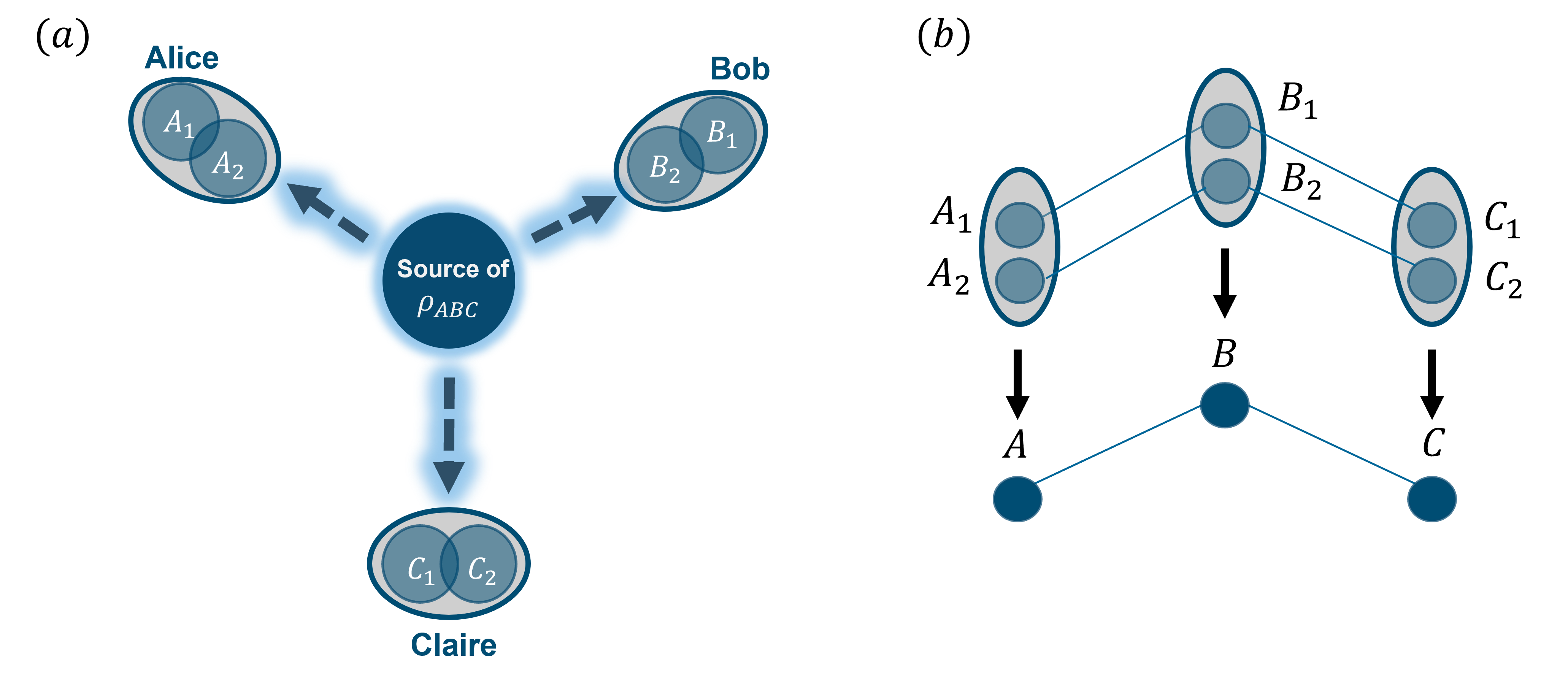}
        \caption{\justifying Superactivation and compression of genuine multipartite entanglement (GME): 
        (a) Superactivation of GME: The three parties share two copies of a biseparable state $\varrho^{ABC}$. If this state is entangled across each bipartition, then the two-copy state $\varrho^{A_1B_1C_1}\otimes \varrho^{A_2B_2C_2}$ might be GME with respect to the partition $A_1A_2|B_1B_2|C_1C_2$, leading to GME superactivation.
        (b) Entanglement compression: Each party applies a local projection to a two-copy state, projecting their two-copy system to the single-copy space again. Since local projections can never create entanglement, the two-copy state must be \textit{at least} as entangled as the resulting state.
        }
        \label{fig:activationAndCompression}
    \end{figure}

    Genuine multipartite entanglement (GME) is a resource that plays an important role in quantum technologies, such as quantum metrology~\cite{Toth2012, HyllusLaskowskiKrischekEtal2012}, or quantum secret sharing~\cite{HilleryBuzekBerthiaume1999,SchauerHuberHiesmayr2010}, and genuinely multipartite entangled states can be used as a benchmark in experiments, as they certify entanglement shared among all parties~\cite{CaoEtAl2023}. 
    Recent studies of multi-copy scenarios in multipartite entanglement theory have shown theoretically~\cite{YamasakiMorelliMiethlingerBavarescoFriisHuber2022, PalazuelosDeVicente2022, BaksovaLeskovjanovaMistaAgudeloFriis2024} and experimentally~\cite{ZhangFeiLiuetal2026, StarekGollerthanLeskovjanovaMethTirlerFriisRingbauerMista2026} that GME can be superactivated from multiple copies of biseparable states. Superactivation describes the phenomenon in which the combination of multiple copies of a state exhibits a property not present in a single copy. For GME, this effect occurs for every biseparable state which is entangled across every bipartition~\cite{PalazuelosDeVicente2022}, see also Fig.~\ref{fig:activationAndCompression}(a). One of the first demonstrations of GME superactivation used a simple projection scheme: All parties apply local projections from the multi-copy space to the single-copy space, thereby certifying in the lower-dimensional space that the projected state is genuinely multipartite entangled. One can then conclude that also the multi-copy state was genuinely multipartite entangled, as local operations cannot create entanglement, see also Fig.~\ref{fig:activationAndCompression}(b).
    
    Since joint operations on multiple copies are experimentally challenging, a natural follow-up question is whether the superactivated GME can be projected back into the single-copy subspace, a property known as \textit{compressibility}. It was shown in Ref.~\cite{WeinbrennerBaksovaDenkerMorelliYuFriisGuhne2024} that this is not always possible: There exist two-copy activatable states whose superactivated GME can never be preserved under any local projection to the single-copy subspace, a phenomenon termed incompressible entanglement (ICE). This motivates us to investigate GME compressibility more broadly, going beyond the two-copy regime studied in Ref.~\cite{WeinbrennerBaksovaDenkerMorelliYuFriisGuhne2024}.\\
    We consider here a fixed projection map and examine the behaviour of entanglement compression in the limit of many copies, demonstrating clear differences between distillation and compression schemes.
    We focus on the Hadamard map as an example, and consider the classes of GHZ-diagonal~\cite{GuehneSeevinck2010} and GHZ-symmetric three-qubit states~\cite{DuerCirac2001, EltschkaSiewert2012,EltschkaSiewert_2012b,EltschkaSiewert_2012c}, where the two-copy case is already well understood~\cite{WeinbrennerBaksovaDenkerMorelliYuFriisGuhne2024}. We find that although almost all considered states exhibit superactivated GME that can be compressed with the Hadamard map, the quality of the compressed GME, quantified by fidelity with pure GHZ states, begins to decrease rapidly once the optimal number of copies for compressing the given state is exceeded. 
    Finally, we use tomographic data from the experimental demonstration of GME superactivation from Ref.~\cite{ZhangFeiLiuetal2026} to assess how sensitive the quality of GME compression under the Hadamard map is to experimental imperfections, comparing the theoretical predictions with the reconstructed states.

\section{Superactivation and compression of genuine multipartite entanglement}

\begin{figure}[t]
    \centering
    \includegraphics[width=0.9\linewidth]{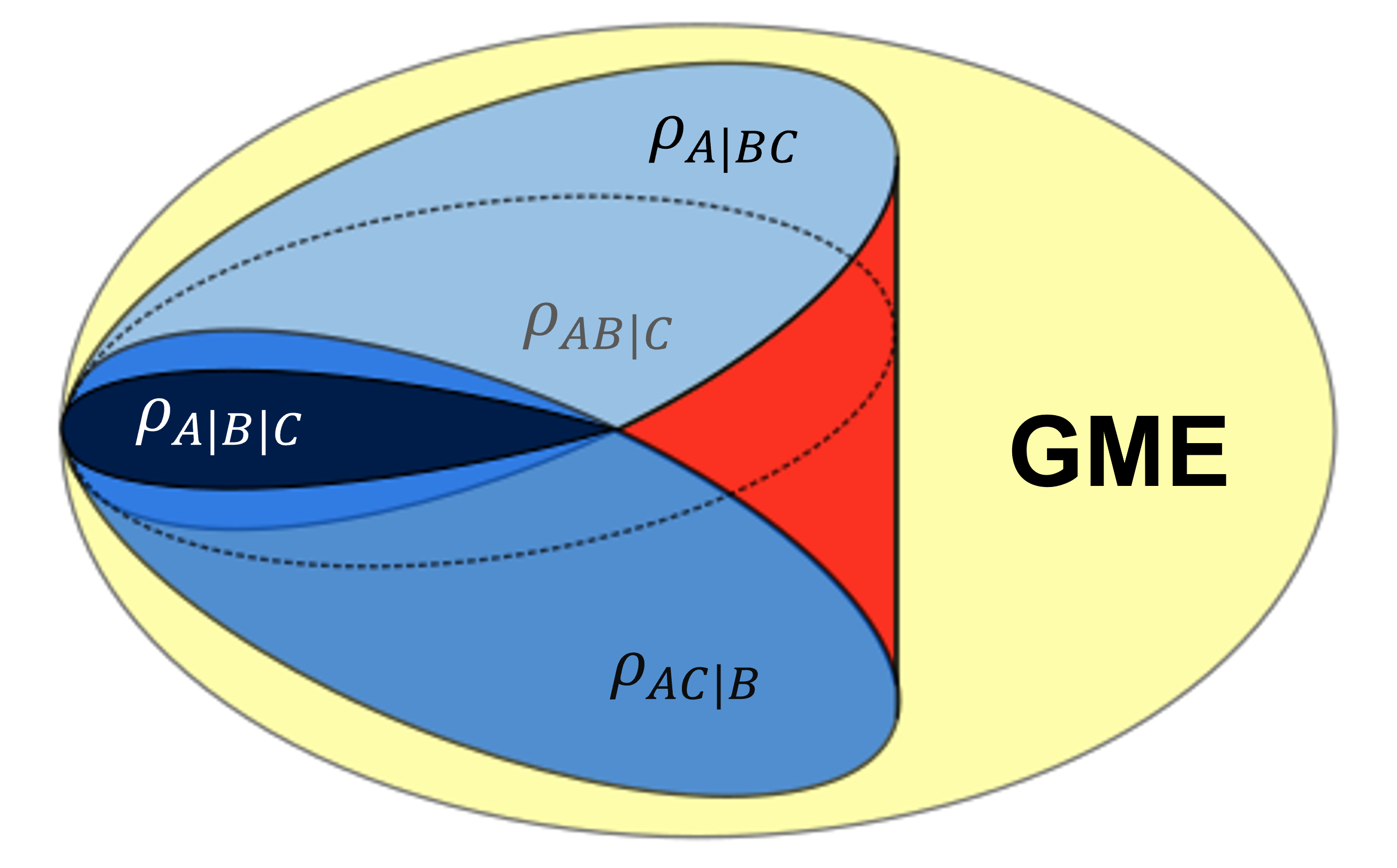}
    \caption{\justifying Separability structure for tripartite systems: 
    Fully separable states $\varrho_{A|B|C}$ (dark blue) form a convex subset of the intersection of the three convex sets of partition-separable states, $\varrho_{AB|C},$ $\varrho_{AC|B}$, and $\varrho_{BC|A}$ (two light-blue regions and background).
    The convex hull of all partition-separable states forms the set of biseparable states (all blue and red regions). 
    The FIB states lie in the red region. 
    All other fully inseparable states that lie outside the set of biseparable states (yellow) are genuinely multipartite entangled.}
    \label{fig:overview}
\end{figure}

    We start by briefly recalling the relevant concepts of multipartite entanglement \cite{GuehneToth2009, FriisVitaglianoMalikHuber2019}. We restrict the discussion to three parties, Alice, Bob, and Claire, but the extension to $N$ parties is straightforward. For a graphical representation of the state space in the tripartite scenario, see also Fig.~\ref{fig:overview}.

    A tripartite state $\varrho^{ABC}$ is called \textit{fully separable} if it admits a decomposition of the form
    \begin{align}
        \varrho^\mathrm{fs} = \sum_i p_i \, \varrho_i^A \otimes \varrho_i^B \otimes \varrho_i^C,
    \end{align}
    with the probability weights $p_i$ satisfying $0\leq p_i \leq 1$ and $\sum_ip_i=1$.
    States that are separable with respect to a fixed partition, for instance $A|BC$,
    \begin{align}
        \varrho^\mathrm{ps} = \sum_i p_i \, \varrho_i^A \otimes \varrho_i^{BC},
    \end{align}
    are called \textit{partition separable}. States which can be expressed as a convex combination of partition-separable states, i.e.
    \begin{align}\label{eq:bisepDecomp}
        \varrho^\mathrm{bs} = p_A\, \varrho^{A|BC} + p_B\, \varrho^{B|AC} + p_C\, \varrho^{C|AB},
    \end{align}
    with $p_A + p_B + p_C = 1$ and $p_A, p_B, p_C \geq 0$, are called \textit{biseparable}. Every remaining state that does not allow for such a decomposition is called \textit{genuinely multipartite entangled}. 
    Lastly, we introduce the notion of \textit{fully inseparable} states, which are states that are not separable for any fixed partition.  In the following considerations, the states which are biseparable, i.e. which admit a decomposition as in Eq.~(\ref{eq:bisepDecomp}), but are not separable for any fixed partition, play an important role. We therefore refer to these states, which are entangled for every bipartition, as \textit{fully inseparable biseparable} (FIB). 
    
    We now consider multiple copies of a given state. For example, suppose that the three parties Alice, Bob and Claire share two copies of a state $\varrho^{ABC}$, that is
    \begin{align}\label{eq:TwoCopyState}
        \varrho_{ABC} = \varrho^{A_1 B_1 C_1} \otimes \varrho^{A_2 B_2 C_2}.
    \end{align}
    Here, the labels $A$, $B$, and $C$ indicate the grouping of subsystems across copies, i.e. $A = A_1 A_2$, $B = B_1 B_2$, and $C = C_1 C_2$, see also Fig.~\ref{fig:activationAndCompression}(a).
    It is straightforward to see that if the single-copy state $\varrho^{A_i B_i C_i}$ is separable with respect to a fixed partition, e.g. $A_1|B_1C_1$, then the tensor product state $\varrho_{ABC}$ is separable with respect to the corresponding grouping, e.g. here $A|BC$. However, if the state is biseparable but not partition-separable for any fixed partition, i.e., FIB, it is not immediately clear that the multi-copy state will again be biseparable due to the resulting cross-terms. 
    Indeed, the multi-copy state may become genuinely multipartite entangled already for two copies~\cite{HuberPlesch2011, YamasakiMorelliMiethlingerBavarescoFriisHuber2022, WeinbrennerBaksovaDenkerMorelliYuFriisGuhne2024}, and it was proven that every FIB state will become genuinely multipartite entangled for a sufficiently large number of copies $k$~\cite{PalazuelosDeVicente2022, BaksovaLeskovjanovaMistaAgudeloFriis2024}. This phenomenon of obtaining a multi-copy GME state from biseparable single-copy states is known as \textit{superactivation of GME}. 

    Moreover, the superactivated GME may be projected back to the single-copy space by local projections, see also Fig.~\ref{fig:activationAndCompression}(b). As local projections cannot create GME, one may use this method to detect the superactivated GME by showing that the projected state is genuinely multipartite entangled, compressing the multi-copy GME to the single-copy level~\cite{YamasakiMorelliMiethlingerBavarescoFriisHuber2022,WeinbrennerBaksovaDenkerMorelliYuFriisGuhne2024}.
    One such map for projecting a multi-copy state onto the single-copy subspace is the \textit{Hadamard map}, which was implemented already in Ref.~\cite{YamasakiMorelliMiethlingerBavarescoFriisHuber2022} to detect superactivation. For a two-copy three-qubit state as in Eq.~(\ref{eq:TwoCopyState}), it takes the form
    \begin{equation}
        \mathcal{E}(\varrho_{ABC})
        \;=\;
        \bigl[E_A\otimes E_B\otimes E_C\bigr]\,
        \varrho_{ABC}\,
        \bigl[E_A^\dagger \otimes E_B^\dagger \otimes E_C^\dagger\bigr],
    \end{equation}
    with
    \begin{equation}
    \label{eq:HadamardKraus}
        E_X \;=\; \ketbra{0_X}{0_{X_1}0_{X_2}} \;+\; \ketbra{1_X}{1_{X_1}1_{X_2}},
    \end{equation}
    for $X=A,B,C$, where each of $A$, $B$, $C$ consists of two subsystems $X_1$ and $X_2$. Note that the output state $\mathcal{E}(\varrho_{ABC})$ may require normalisation. When acting locally on each subsystem, this map applied to a tensor product $\varrho_1\otimes\varrho_2$ of two arbitrary input states (in our case, two copies of $\varrho^{ABC}$) admits the compact expression~\cite{LamiHuber2016, HolmesCobleSornborgerSuba2023}
    \begin{equation}
        \mathcal{E}(\varrho_1\otimes\varrho_2) \;=\; \varrho_1\circ\varrho_2,
    \end{equation}
    where ``$\circ$'' denotes the Schur (or Hadamard) product, i.e. the component-wise product of matrices. The map can be implemented on two states by an application of a CNOT gate followed by a $\ketbra{0}{0}$ measurement on the control qubit~\cite{HolmesCobleSornborgerSuba2023}, and an experimental demonstration of GME superactivation and compression via this map was performed in Ref.~\cite{ZhangFeiLiuetal2026}. One should note that this is a probabilistic scheme, which yields the desired outcome with probability $p_\mathrm{succ}=\tr(\varrho\circ \varrho)$. A single-shot enhancement of bipartite entanglement, implemented via a CNOT gate combined with postselection applied to a product of two states encoded in two degrees of freedom of a single photon pair, was realised experimentally in Ref.~\cite{EckerSohrBullaHuberBohmannUrsin2021}.

    The application of the Hadamard map extends naturally to multiple copies of a given state. By associativity and scalar homogeneity of the Schur product, compressing $k$ identical copies of $\varrho$ yields the compact formula
    \begin{equation}
    \label{eq:HadamardMap}
        \tilde{\varrho}_k
        \;=\;
        \frac{1}{\mathcal{N}}
        \mathcal{E}_{k-1}\!\left(\varrho^{\otimes k}\right)
        \;=\;
        \frac{\varrho^{\circ k}}{\operatorname{Tr}\!\left(\varrho^{\circ k}\right)},
    \end{equation}
    where $\mathcal{N}$ denotes a normalization constant. The expression in Eq.~(\ref{eq:HadamardMap}) is particularly convenient for studying GME properties in the many-copy regime: Rather than tracking an exponentially growing Hilbert space dimension, the entanglement properties of $\tilde{\varrho}_k$ can be studied directly through the element-wise $k$-th powers of the single-copy density matrix, for which analytical criteria are often available.

    Obviously, the Hadamard map will not map every superactivated multi-copy state back to a single-copy GME state. Naturally, for arbitrary states, more general maps might be needed for entanglement compression, but there are states where, for a fixed number of copies, the scheme of entanglement compression will never lead to single-copy GME~\cite{WeinbrennerBaksovaDenkerMorelliYuFriisGuhne2024}.

\section{Entanglement compression by Hadamard map of GHZ-diagonal states}\label{Sec:GHZDiagonal}

    We now want to investigate the behaviour of arbitrary states under the compression scheme in which the number of copies $k$ increases, for a fixed projection map. Fixing the map to the Hadamard map discussed above, we first focus on GHZ-diagonal states, since they remain GHZ-diagonal under compression with the Hadamard map. As the name suggests, a GHZ-diagonal state is diagonal in the GHZ basis, formed by the eight vectors $\{ (\ket{ijk} \pm \ket{\bar{i}\bar{j}\bar{k}})/\sqrt{2}\}_{i,j,k=0,1}$ where $\ket{\bar{b}}$ denotes the bit-flipped value of the qubit $\ket{{b}}$. In the computational basis, these states are in X-form, having only (nonnegative) entries $\lambda_i$ for $i=1,2,3,4$ on the diagonal, and $\mu_i$ for $i=1,2,3,4$ on the antidiagonal, i.e.
    \begin{equation}\label{eq:xform}
        \varrho_X=\frac{1}{2\sum_{i=1}^4\lambda_i}\begin{pmatrix}
        \lambda_1 & 0 & 0 & 0 & 0 & 0 & 0 & \mu_1 \\
        0 & \lambda_2 & 0 & 0 & 0 & 0 & \mu_2 & 0 \\
        0 & 0 & \lambda_3 & 0 & 0 & \mu_3 & 0 & 0 \\
        0 & 0 & 0 & \lambda_4 & \mu_4 & 0 & 0 & 0 \\
        0 & 0 & 0 & \mu_4 & \lambda_4 & 0 & 0 & 0 \\
        0 & 0 & \mu_3 & 0 & 0 & \lambda_3 & 0 & 0 \\
        0 & \mu_2 & 0 & 0 & 0 & 0 & \lambda_2 & 0 \\
        \mu_1 & 0 & 0 & 0 & 0 & 0 & 0 & \lambda_1
    \end{pmatrix}.
    \end{equation}
    In order to investigate entanglement, we may assume that the entries $\lambda_i$ are in decreasing order, i.e. $\lambda_1 \geq \lambda_2 \geq \lambda_3 \geq \lambda_4 \geq 0$. Additionally, the off-diagonal terms need to fulfil $|\mu_i| \leq \lambda_i$ for the density matrix to be positive. Then, every such state is biseparable if and only if~\cite{GuehneSeevinck2010}
    \begin{equation}\label{eq:bisepcrit}
        |\mu_1| \leq \lambda_2 + \lambda_3 + \lambda_4.
    \end{equation}
    Following the investigations in Ref.~\cite{WeinbrennerBaksovaDenkerMorelliYuFriisGuhne2024}, we restrict these states even further by considering only the states where $\mu_i=\lambda_i$, using the short-hand notation $\varrho=\chi(\lambda_1,\lambda_2,\lambda_3,\lambda_4)$, where the normalisation is implicitly assumed.
    It is straightforward to see that $k$ copies of the state $\chi(\lambda_1,\lambda_2,\lambda_3,\lambda_4)$ will be mapped by the Hadamard map to the GHZ-diagonal state $\chi(\lambda_1^k,\lambda_2^k,\lambda_3^k,\lambda_4^k)$. Applying the criterion in Eq.~\eqref{eq:bisepcrit}, one therefore finds that $\chi(\lambda_1,\lambda_2,\lambda_3,\lambda_4)$ is $k$-copy activatable, if $\lambda_1^k > \lambda_2^k + \lambda_3^k + \lambda_4^k$.

\subsection{Fixed points}

    We start by characterizing those states $\chi(\lambda_1,\lambda_2,\lambda_3,\lambda_4)$ that are fixed points of the Hadamard map. Since this map can be applied iteratively, it suffices to study the single-step map for stability analysis.

    We are therefore searching for states that satisfy $\varrho=\frac{\varrho\circ\varrho}{\tr(\varrho\circ\varrho)}$. Denoting the normalization as $\tau:=\tr(\varrho\circ\varrho)>0$, this implies that all non-zero matrix elements of $\varrho$ satisfy $\varrho_{ij}=\tau$. Setting $s$ to be the number of non-zero $\lambda_i$ of $\chi(\lambda_1,\lambda_2,\lambda_3,\lambda_4)$, we find $\tau=\frac{1}{2s}$. Noting that the $\lambda_i$ are ordered, one finds that the states satisfying this condition are exactly given by $\chi(1,0,0,0)$, $\chi(1,1,0,0)$, $\chi(1,1,1,0)$, and $\chi(1,1,1,1)$. These are the four equally weighted convex mixtures of the GHZ states $\{ (\ket{ijk} + \ket{\bar{i}\bar{j}\bar{k}})/\sqrt{2}\}_{i,j,k=0,1}$ containing one to four different states.

    In the one-copy case, the criterion in Eq.~(\ref{eq:bisepcrit}) directly detects $\chi(1,0,0,0)$, i.e. the pure GHZ state, as entangled. Further, the fixed point $\chi(1,1,0,0)$ is partition-separable with respect to the partition $AB|C$, the point $\chi(1,1,1,0)$ is FIB, and finally, the state $\chi(1,1,1,1)$ is fully separable.

\begin{figure}
        \includegraphics[width=0.75\linewidth]{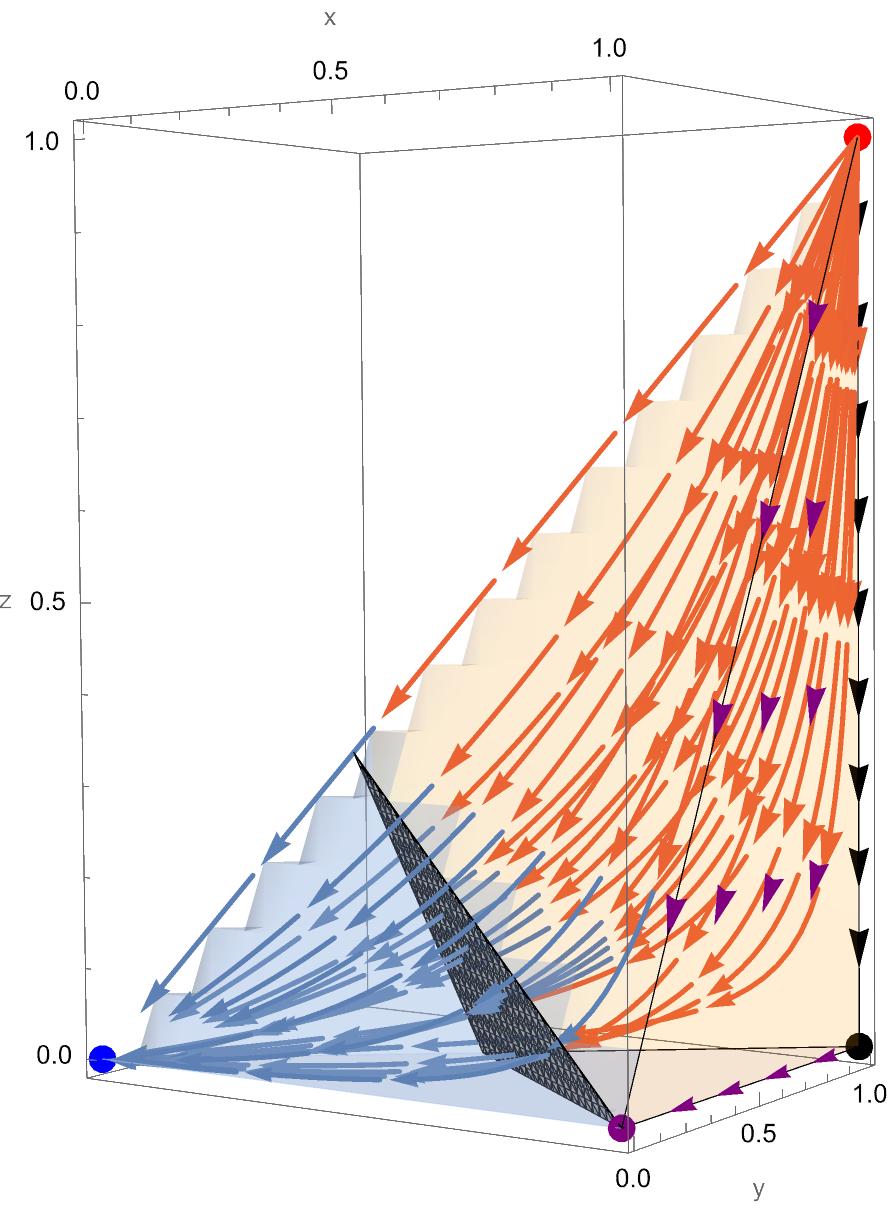}
        \caption{\justifying Entanglement compression of a large number of copies under the Hadamard map for states $\chi(1,x,y,z)$: The red region corresponds to biseparable states, while the blue region depicts genuinely multipartite entangled states. The boundary between these regions is given by the grey plane. The red dot is the repellor and corresponds to the fully separable state $\chi(1,1,1,1)$, the blue point is the attractor and corresponds to the pure GHZ state $\chi(1,0,0,0)$. The purple point is the partition-separable saddle point $\chi(1,1,0,0)$, and the black point is the FIB saddle point $\chi(1,1,1,0)$. 
        One can see that all the states in the interior of the polytope converge to the pure GHZ state. 
        In contrast, the points on the boundary $x=1$ defined by a 2-dimensional plane converge to the two saddle points. The subset on the 1-dimensional line given by $x=1$ and $y=1$ converges to the point $\chi(1,1,1,0)$ (black arrows), while the rest of the plane converges to the saddle point $\chi(1,1,0,0)$ (purple arrows). All points in this plane correspond to states which can never be compressed with the Hadamard map, even in the limit of infinitely many copies.
        }
        \label{fig:3dvectorflow}
    \end{figure}

    We now choose, without loss of generality, the parametrisation $\chi(1,x,y,z)$, which enables graphical depictions of the state space. The set of biseparable states is then a polytope in 3-dimensional space, depicted in Fig.~\ref{fig:3dvectorflow}. We note that one application of the Hadamard map to two copies of the state $\chi(1,x,y,z)$ leads to the updated state $\chi(1,\Tilde{x},\Tilde{y},\Tilde{z})=\chi(1,x^2,y^2,z^2)$, which can be interpreted as a map in a 3-dimensional real space. One may therefore define a vector field that is approximated via the Euler method by the Hadamard map to investigate the stability of the Hadamard map (see Appendix~\ref{app-sec-EulerMethod} and, e.g.~\cite[Chapter 2.8]{Strogatz1994})  by considering
    \begin{equation}
        \mathbf{F}(x,y,z)=\begin{pmatrix}
            x^2-x\\
            y^2-y\\
            z^2-z
        \end{pmatrix}.
    \end{equation}
    The vector flow of this field is shown in Fig.~\ref{fig:3dvectorflow}. For any state from the interior of the polytope satisfying $\chi(1, x, y, z)$, $\{x, y, z\} \in (0, 1)$, there always exists a finite $k$ for which the compressed state becomes genuinely multipartite entangled. Noting that $\max\{x, y, z\} = x$, the state is genuinely multipartite entangled whenever
    \begin{equation}
        k > \left\lceil \frac{\ln(1/3)}{\ln (x)} \right\rceil,
    \end{equation}
    which follows from $1 > 3x^k \geq x^k + y^k + z^k$, and also provides an upper bound on the minimal sufficient number of copies. In the asymptotic limit $k \to \infty$, all interior states converge to the pure GHZ state $\chi(1, 0, 0, 0)$, since $x, y, z \in (0,1)$ implies
    \begin{equation}
        \chi(1^k, x^k, y^k, z^k) \xrightarrow{k \to \infty} 
        \chi(1, 0, 0, 0).
    \end{equation}
    Hence, in terms of stability, the genuinely multipartite entangled state is an attractor, the partition-separable state and the FIB state are saddle points, and the fully separable state is a repellor~\cite{Strogatz1994}.

    One can also see in Fig.~\ref{fig:3dvectorflow} that all states on the right boundary of the polytope can never be compressed by the Hadamard map, not even in the limit of infinitely many copies, as they converge to one of the saddle points. However, some of these states are compressible under more general maps~\cite{WeinbrennerBaksovaDenkerMorelliYuFriisGuhne2024}. For example, it was shown that the state $\chi(1,1,0.05,0)$ achieves a high fidelity of 0.97561 with the GHZ state already after compressing only two copies with an optimised map. This example highlights that the usefulness of a given state in the compression scheme depends heavily on the local projections available in a particular experimental implementation.

\subsection{Optimal number of copies}

    We have seen that, in principle, every state in the interior of the polytope may be compressed with the Hadamard map while preserving GME. However, this leaves the question open of \textit{how good} the compressed entanglement is. As a quantifier, we consider here the fidelity with the pure GHZ state
    \begin{align}
        \ket{\mathrm{GHZ}_+} = \frac{1}{\sqrt{2}}(\ket{000} + \ket{111}).
    \end{align}
    One can easily check that the fidelity of the state $\chi(1,x,y,z)$ after compressing $k$ copies with the Hadamard map is given by
    \begin{equation}\label{eq:fidelity}
        \mathcal{F}(\mathrm{GHZ},\chi(1,x,y,z)^{\circ k})=\frac{1}{1+x^k+y^k+z^k}.
    \end{equation}
      
    Hence, for the purpose of achieving a fidelity with the pure GHZ state larger than $1-\varepsilon$, one can define a sufficient number of copies to achieve such fidelity, i.e. $k$ such that $\mathcal{F}(\mathrm{GHZ},\chi(1,x,y,z)^{\circ k}) \geq 1-\varepsilon$. Noting that  $\max\{x,y,z\} = x$, an upper bound on the minimal number of copies required is given by
    \begin{equation}\label{eq:fidelitybound}
          k>\frac{\ln\left[\frac{\varepsilon}{3\left(1-\varepsilon\right)}\right]}{\ln(x)}.
    \end{equation}
    This implies that in the compression scheme using the Hadamard map, every interior point of the polytope can achieve arbitrarily high GHZ fidelity with sufficiently many copies.
    
\begin{figure*}[!tbp]
  \begin{subfigure}[b]{0.275\textwidth}
    \includegraphics[width=\textwidth]{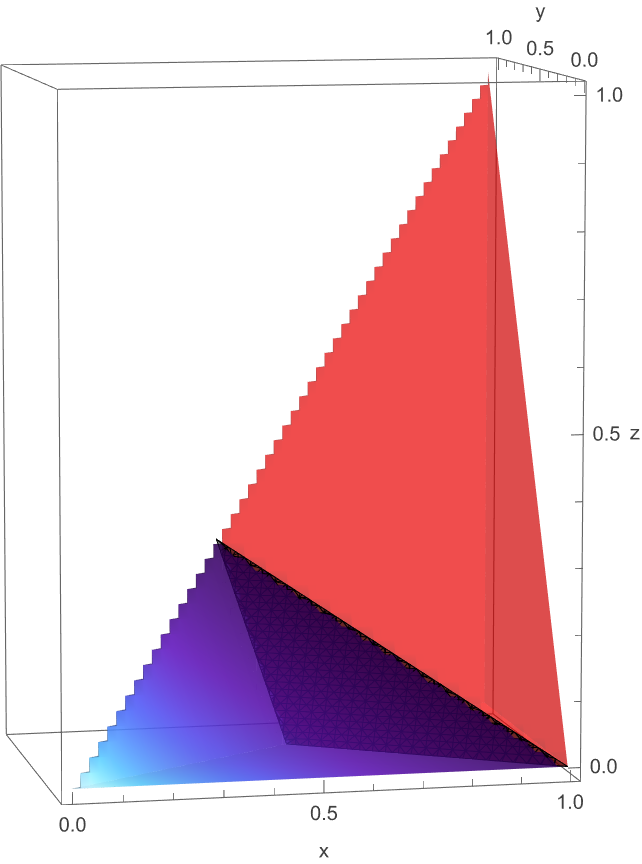}
    \caption{$p=0, k=1$}
  \end{subfigure}
  \hfill
  \begin{subfigure}[b]{0.275\textwidth}
    \includegraphics[width=\textwidth]{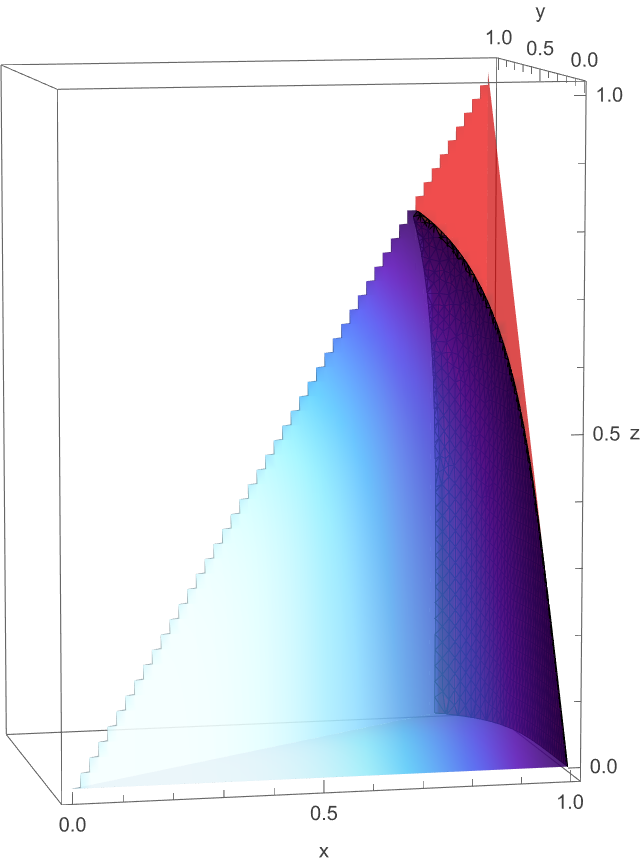}
    \caption{$p=0, k=5$}
  \end{subfigure}
   \hfill
  \begin{subfigure}[b]{0.275\textwidth}
    \includegraphics[width=\textwidth]{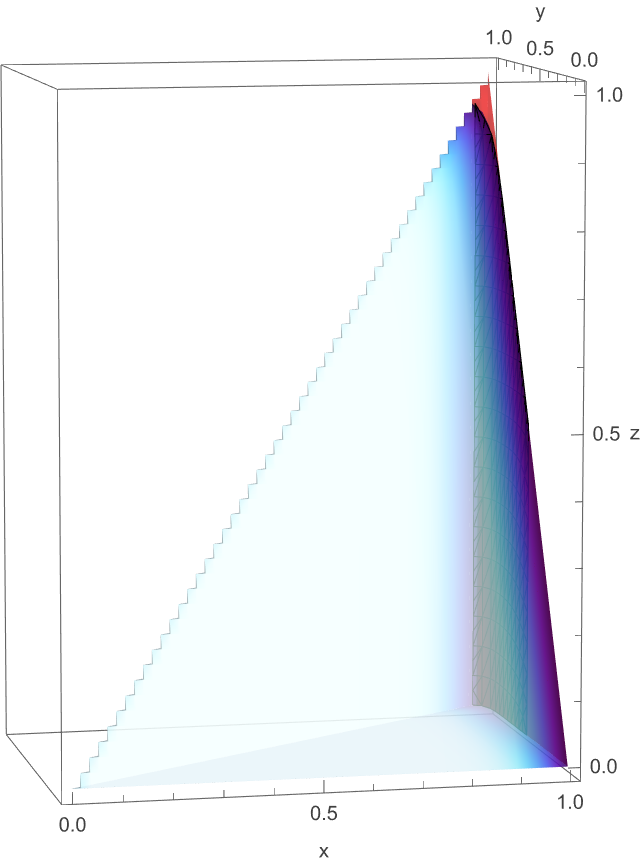}
    \caption{$p=0, k=25$}
  \end{subfigure}
   \hfill
  \begin{subfigure}[b]{0.275\textwidth}
    \includegraphics[width=\textwidth]{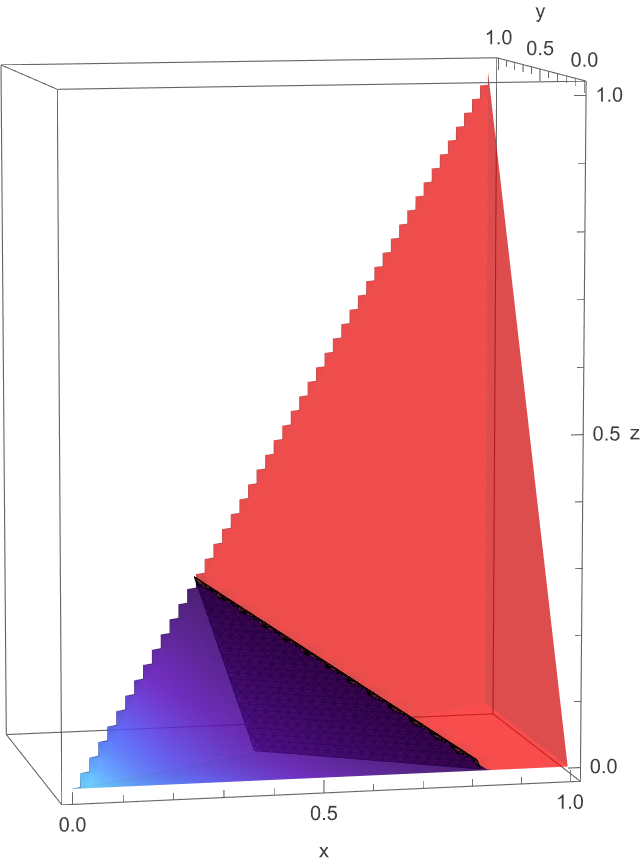}
    \caption{$p=0.1, k=1$}
  \end{subfigure}
  \hfill
  \begin{subfigure}[b]{0.275\textwidth}
    \includegraphics[width=\textwidth]{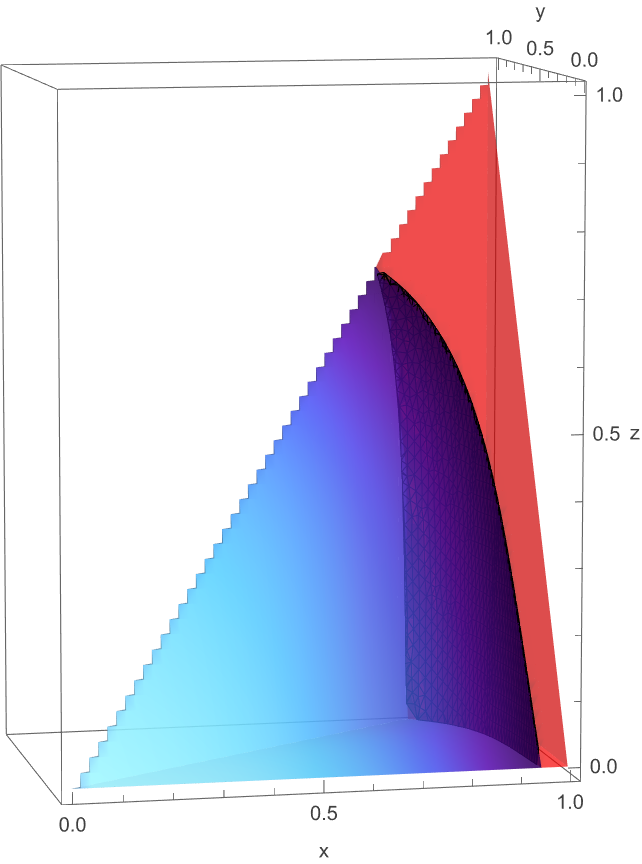}
    \caption{$p=0.1, k=5$}
  \end{subfigure}
  \hfill
  \begin{subfigure}[b]{0.275\textwidth}
    \includegraphics[width=\textwidth]{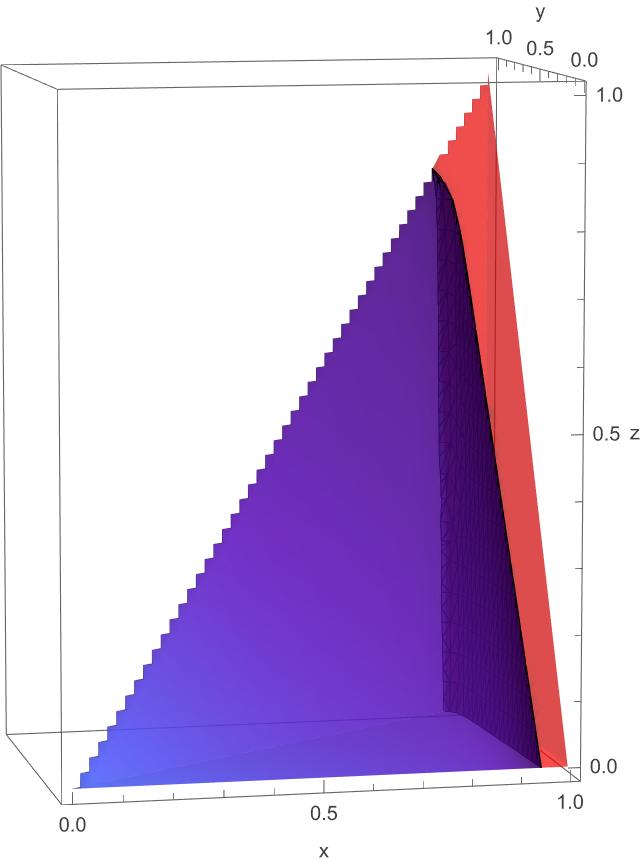}
    \caption{$p=0.1, k=25$}
  \end{subfigure}
  \hfill
  \begin{subfigure}[b]{0.4\textwidth}
    \includegraphics[width=\textwidth]{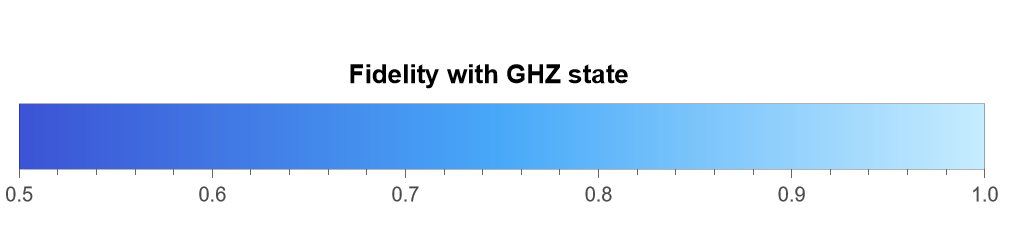}
  \end{subfigure}
  \caption{\justifying Entanglement compression under the Hadamard map for GHZ-diagonal states $\chi(1,x,y,z)$ without white noise vs. with white noise: The plots show GME regions (blue) vs. biseparable regions (red) of states $(1-p)\chi(1,x,y,z)+\frac{p}{8}\mathbb{1}_8$ for added white noise $p=0$ and $p=0.1$, and for $k=1,5,25$ copies of the state.  
  In the absence of added white noise (upper row), one finds that the GME region and the fidelity of the compressed state with the GHZ state increase as the number of copies grows. On the other hand, with added white noise (lower row), the GME region still increases with the number of copies compressed, whereas fidelity increases only up to a certain number of copies and then decreases.
  }
  \label{fig:NoisyGHZDiagonal}
\end{figure*}

    However, the situation significantly changes in the presence of white noise, i.e. if we consider states of the form $(1-p)\chi(1,x,y,z)+\frac{p}{8}\mathbb{1}_8$. In Fig.~\ref{fig:NoisyGHZDiagonal}, we show in comparison the behaviour of states without and with added white noise with noise parameter $p=0.1$ under entanglement compression by the Hadamard map. There, the red region depicts states which are mapped to biseparable states in the compression scheme, while states in the blue region are mapped to genuinely multipartite entangled states. The different shades of blue correspond to different values of fidelity with the pure GHZ state. One can see that for the states $(1-p)\chi(1,x,y,z)+\frac{p}{8}\mathbb{1}_8$ with $p=0$ (upper row), all states, except the boundary plane, become genuinely multipartite entangled with increasing number of copies, while the fidelity with the pure GHZ state also increases, as noted above. This coincides with the convergence of the vector field shown in Fig.~\ref{fig:3dvectorflow}. For the states $(1-p)\chi(1,x,y,z)+\frac{p}{8}\mathbb{1}_8$ with $p=0.1$ (bottom row), the size of the GME region also increases for an increasing number of copies. However, the fidelity of the compressed states with the pure GHZ states first increases for $k=5$ but then decreases for $k=25$.

    Actually, for any noise parameter $p>0$, we find that the largest diagonal element is greater than all off-diagonal elements, i.e. $\lambda_{i_\mathrm{max}}>\mu_{i_\mathrm{max}}$. Since the compression with the Hadamard map takes these elements to the $k$-th power, only the largest diagonal element(s) of the density matrix will not vanish in the asymptotic limit of compressing infinitely many copies, while all off-diagonal elements will vanish. This implies that all GHZ-diagonal states with added white noise will converge to fully separable states. 

    This claim can be generalised for any state with added white noise, as the Hadamard map under $k\to\infty$ allows only the largest density-matrix elements in the given basis to survive. Since any density matrix must be positive semidefinite, meaning the off-diagonal elements are upper-bounded by diagonal elements, any addition to the diagonal elements will make the off-diagonal elements vanish in the infinite limit, and thus the state will converge to a fully separable state. 

    Vaguely speaking, one could conclude that, as the number of copies compressed by the Hadamard map increases, the GME becomes more ``fragile" for a fixed state. We will demonstrate this behaviour using the state $\chi(5,4,3,0)$. The state is two-copy activatable and also two-copy compressible~\cite{WeinbrennerBaksovaDenkerMorelliYuFriisGuhne2024} with an optimised projection map, but not two-copy compressible with the Hadamard map. However, we find that it is compressible with the Hadamard map for three copies.
    Analysing the GHZ fidelity of this state with added white noise, i.e. $\varrho(p) =(1-p)\chi(5,4,3,0)+\frac{p}{8}\mathbb{1}_8$, we find again that the fidelity first increases and then decreases for a higher number of copies $k$. This implies that for a given state, there is an ideal number of copies, $k$, that one should compress to achieve the highest possible GHZ fidelity. The results for the GHZ fidelity are depicted in Fig.~\ref{fig:chi(5,4,3,0)}, where the black curve marks the ideal number of copies for a given value of $p$.

    \begin{figure}[!t]
        \includegraphics[width=0.8\linewidth]{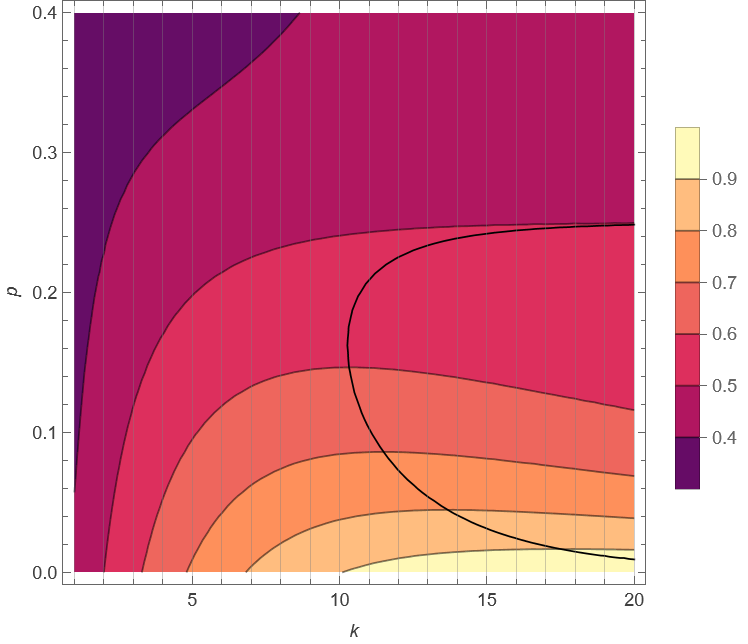}
        \caption{\justifying GHZ fidelity after compressing $k=1,\dots, 20$ copies of $(1-p)\chi(5,4,3,0)+p\mathbb{1}_8/8$ with the Hadamard map. Combinations of $p$ and $k$ where the GHZ fidelity is equal to or greater than the value depicted by the heat map. 
        The black line denotes the ideal number of copies one should consider for a given noise parameter $p$ to maximise the GHZ fidelity. The minimal ideal number of copies is reached for the noise parameter $p^*\approx 0.16$, 
        where the ideal number of copies is given by $k^*\approx 10$.
        }
    \label{fig:chi(5,4,3,0)}
    \end{figure}

\section{Entanglement compression under the Hadamard map of GHZ-symmetric states}

    In this section, we will investigate compression under the Hadamard map for another subclass of GHZ-diagonal states, called GHZ-symmetric states~\cite{DuerCirac2001,EltschkaSiewert_2012b, EltschkaSiewert_2012c, EltschkaSiewert2012}. These states can be described by convex combinations of the states $\ket{\mathrm{GHZ}_\pm} = (\ket{000} \pm \ket{111})/\sqrt{2}$ and $\varrho_{\mathrm{sym}}=(\mathbb{1}_8-\ketbra{000}{000}-\ketbra{111}{111})/6$. Conveniently, one can depict all GHZ-symmetric states on a two-dimensional plane defined by coordinates $[x,y]$, where the parametrisation is given by~\cite{EltschkaSiewert2012}
    \begin{align}\label{eq:GHZsymCoordinates}
        x[\varrho]&=(\sum_{m=\pm}m\bra{\mathrm{GHZ}_m}\varrho\ket{\mathrm{GHZ}_m})/2\\
        y[\varrho]&=(-{1}/{4}+\sum_{m=\pm}\bra{\mathrm{GHZ}_m}\varrho\ket{\mathrm{GHZ}_m})/\sqrt{3}
    \end{align}
    For this family of states, the parameter regions of the entanglement classes of full separability, biseparability, W-type entanglement, and GHZ-type entanglement have been fully characterized~\cite{EltschkaSiewert2012}, and in Ref.~\cite{WeinbrennerBaksovaDenkerMorelliYuFriisGuhne2024}, states of this class have been investigated in the context of GME superactivation and compression under the Hadamard map, up to the four-copy case. Here, we provide a full characterisation of GME compression under the Hadamard map of these states, including the asymptotic regimes, and show that the GME can be compressed by the Hadamard map for all GME-activatable GHZ-symmetric states, thereby determining the minimal number of copies needed.

\begin{figure}[t]
    \centering
    \includegraphics[width=1.1\linewidth]{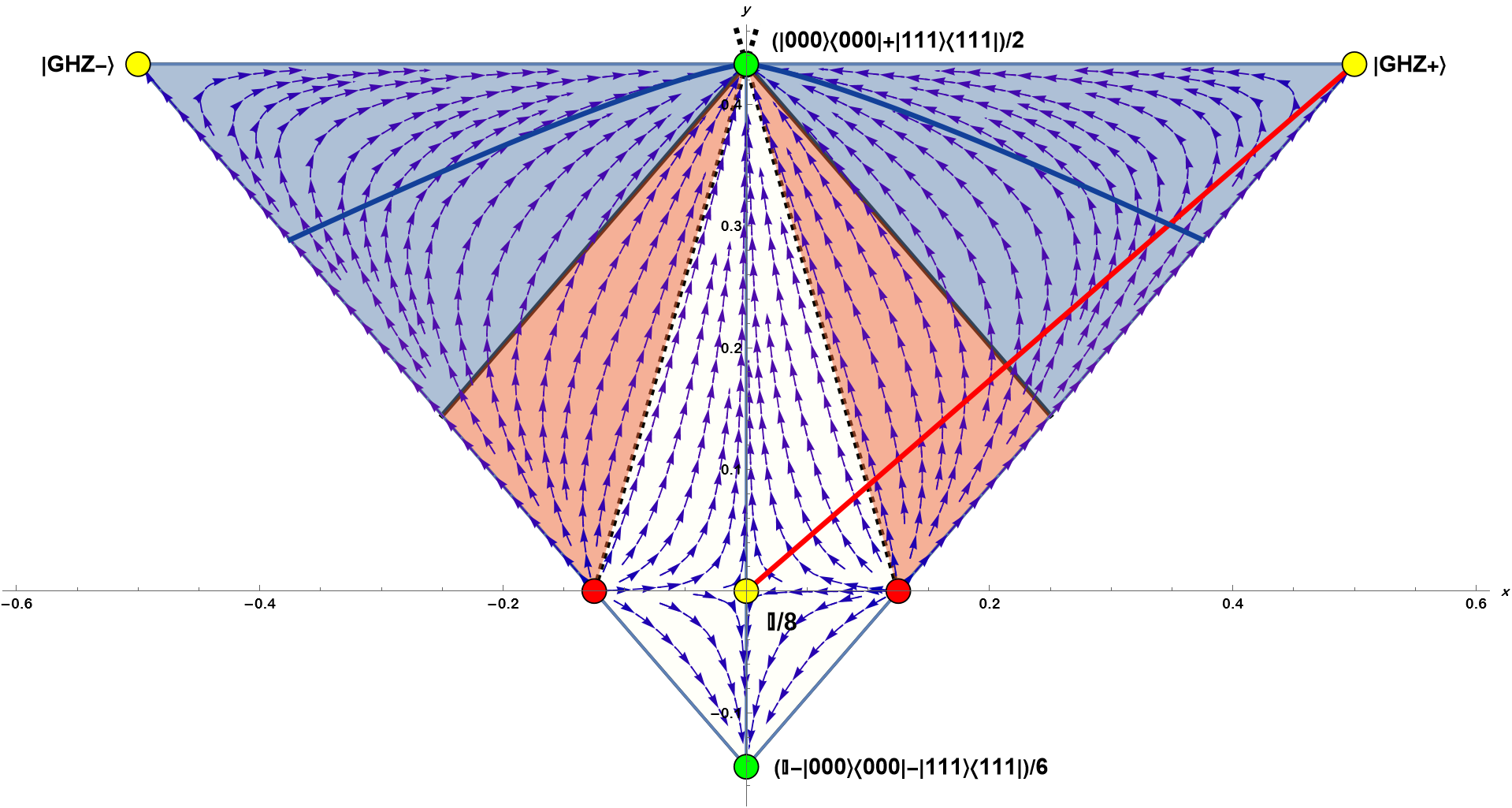}
    \caption{\justifying Flow of the vector field from Eq.~(\ref{eq:vectorfield}). The two green points are attractors corresponding to fully separable states $\left(\ket{000}\!\bra{000}+\ket{111}\!\bra{111}\right)/2$ and $\left(\mathbb{1}_8-\ket{000}\!\bra{000}-\ket{111}\!\bra{111}\right)/6$, the three yellow points are saddle points, where two of them correspond to pure genuinely multipartite entangled states $\ket{\mathrm{GHZ}_+}$ and $\ket{\mathrm{GHZ}_-}$ and the third one corresponds to the maximally mixed state $\mathbb{1}_8/8$, and the two red points are repellors corresponding to the states $\frac{1}{4}\ket{\mathrm{GHZ}_\pm}\!\bra{\mathrm{GHZ}_\pm}+\frac{1}{8}\left(\mathbb{1}_8-\ket{000}\!\bra{000}-\ket{111}\!\bra{111}\right)$. The red regions correspond to FIB states (GME activatable states). The uncoloured region under the FIB states corresponds to fully separable states, and the blue region above the FIB states corresponds to genuinely multipartite entangled states, where the blue curve separates the region of the W SLOCC class (lower part) and the GHZ SLOCC class (upper part).
    On the red line lie the states $(1-p)\ketbra{\mathrm{GHZ}_+}{\mathrm{GHZ}_+} + \frac{p}{8}\mathbb{1}_8$.
    }
    \label{fig:vectorfield}
\end{figure}

    We again start by analysing the stability of the map~(\ref{eq:HadamardMap}). As the Hadamard map maps multiple copies of a GHZ-symmetric state back to a GHZ-symmetric state, we find in the two-copy case that 
    $\varrho(\tilde{x},\tilde{y}) \propto \varrho^{\circ 2}(x,y) $ with the coordinates
    \begin{align}\label{eq:stepmap}
        \Tilde{x}=f_x(x,y)=\frac{8x^2}{1+16y^2}\quad \text{and} \quad \Tilde{y}=f_y(y)=\frac{2(3y+4\sqrt{3}y^2)}{3(1+16y^2)}.
    \end{align}
    As before, this map approximates via the Euler method a vector field given by the system of differential equations
    \begin{align}\label{eq:vectorfield}
        \dot{x}=v_x(x,y)=f_x(x,y)-x\quad \text{and}\quad \dot{y}=v_y(y)=f_y(y)-y.
    \end{align}
    One should note that the map does not approximate the vector field well in the regions where the map steps are large. However, we can still use the vector field for the stability analysis, as the vector field and the map yield the same stability classification near their fixed points (see Appendix~\ref{app-sec-EulerMethod}).
    The resulting vector field~(\ref{eq:vectorfield}) on the simplex of GHZ-symmetric states with its fixed points is shown in Fig.~\ref{fig:vectorfield}. We note that the two halves of the GHZ-symmetric simplex are equivalent up to a local phase flip: for states with $x<0$, the sign of the off-diagonal element is preserved for odd k and reversed for even k, such that the corresponding dynamics can be identified with that for $x>0$ up to this local phase.

    There, one immediately finds that the pure GHZ states $\ket{\mathrm{GHZ}_\pm}$ are saddle points, and that only the subset of the one-dimensional boundary of the triangle converges to them. Therefore, the set of activatable states that converge to the pure genuinely multipartite entangled state under compression with the Hadamard map is measure zero, while all other activatable states converge to
    the attractor $(\ketbra{000}{000} + \ketbra{111}{111})/2$, i.e. to a fully separable state. This shows that entanglement compression can lead to GME for a fixed map, but the resulting GME may degrade under repeated applications of this scheme.
  
    Interestingly, for GHZ-symmetric states, the minimal number of copies required for them to become genuinely multipartite entangled under compression with the Hadamard map can be readily determined, enabling a more detailed study of the activatable states. Note that GHZ-symmetric states are given by 
    \begin{equation}\label{eq:GHZsym}
        \begin{split}
        \varrho_{\mathrm{sym}} &= p_1\ket{\mathrm{GHZ}_+}\!\bra{\mathrm{GHZ}_+} + 
        p_2\ket{\mathrm{GHZ}_-}\!\bra{\mathrm{GHZ}_-}\\
        &+\frac{(1-p_1-p_2)}{6}\left(\mathbb{1}_8-\ket{000}\!\bra{000}-\ket{111}\!\bra{111}\right),
        \end{split}
    \end{equation}
    with $p_1=\bra{\mathrm{GHZ}_+}\varrho\ket{\mathrm{GHZ}_+}$ and $p_2=\bra{\mathrm{GHZ}_-}\varrho\ket{\mathrm{GHZ}_-}$. Therefore, their density matrix is in X-form and given by
    \begin{equation}\label{eq:GHZsymState}
        \varrho_\mathrm{sym}=\begin{pmatrix}
            A & 0 & 0 & 0 & 0 & 0 & 0 & B \\
            0 & C & 0 & 0 & 0 & 0 & 0 & 0 \\
            0 & 0 & C & 0 & 0 & 0 & 0 & 0 \\
            0 & 0 & 0 & C & 0 & 0 & 0 & 0 \\
            0 & 0 & 0 & 0 & C & 0 & 0 & 0 \\
            0 & 0 & 0 & 0 & 0 & C & 0 & 0 \\
            0 & 0 & 0 & 0 & 0 & 0 & C & 0 \\
            B & 0 & 0 & 0 & 0 & 0 & 0 & A \\
        \end{pmatrix},
    \end{equation}
    with
    \begin{equation}
        A=\frac{p_1+p_2}{2},\quad B=\frac{p_1-p_2}{2},\quad \text{and}\quad C=\frac{1-p_1-p_2}{6}.
    \end{equation}
    Compressing $k$ copies of this state with the Hadamard map, we find
    \begin{equation}
       \varrho^{\circ k}_\mathrm{sym}=\frac{1}{\mathcal{N}}\begin{pmatrix}
           A^k & 0 & 0 & 0 & 0 & 0 & 0 & B^k \\
           0 & C^k & 0 & 0 & 0 & 0 & 0 & 0 \\
           0 & 0 & C^k & 0 & 0 & 0 & 0 & 0 \\
           0 & 0 & 0 & C^k & 0 & 0 & 0 & 0 \\
           0 & 0 & 0 & 0 & C^k & 0 & 0 & 0 \\
           0 & 0 & 0 & 0 & 0 & C^k & 0 & 0 \\
           0 & 0 & 0 & 0 & 0 & 0 & C^k & 0 \\
           B^k & 0 & 0 & 0 & 0 & 0 & 0 & A^k \\
       \end{pmatrix},
    \end{equation}
    with $\mathcal{N}=2A^k+6C^k$. 
    Since this is a GHZ-diagonal state, it is genuinely multipartite entangled if and only if \cite{GuehneSeevinck2010}

    \begin{equation}
       k>\frac{\ln(3)}{\ln(|B|/C)}=\frac{\ln(3)}{\ln\left(\frac{3 |p_1-p_2|}{1-p_1-p_2}\right)}
    \end{equation}
    holds whenever the logarithm in the denominator is positive. As we show below, this condition is satisfied for every FIB state, thereby ensuring that there exists a finite number of copies k for which the compressed state is genuinely multipartite entangled. 
    For the special case of the GHZ state with additional white noise, i.e. $\varrho(p) = (1-p) \ketbra{\mathrm{GHZ}}{\mathrm{GHZ}} + \frac{p}{8}\mathbb{1}_8$, which lies on the line between two saddle points, this translates to the state being $k$-copy compressible by the Hadamard map for
    \begin{equation}\label{eq:GHZcompressionBound}
        k>\frac{\ln(3)}{\ln\left[\frac{4(1-p)}{p}\right]}\quad \text{for}\quad p<\frac{4}{5}.
    \end{equation}

    A further observation of Fig.~\ref{fig:vectorfield} can be made regarding which FIB states can be compressed. It is evident that the trajectories of some FIB states exit the FIB region after a finite number of steps $k$, suggesting that compression with the fixed Hadamard map can lead to GME for these states. One may therefore ask whether \textit{all} FIB GHZ-symmetric states are compressible by the Hadamard map. Indeed, we will answer this question in the affirmative.

    \noindent\textbf{Lemma:} \textit{Every FIB GHZ-symmetric state $\varrho_\mathrm{FIB}$ exhibits compressible GME under the Hadamard map if a sufficiently large number of copies $k$ of the state $\varrho_\mathrm{FIB}^{\otimes k}$ is compressed.}

    \noindent\textit{Proof:} We will prove Lemma 1 for the right part of the simplex, which implies that the equivalent proof can be done for the left part, considering mirror symmetry. The line separating the partition-separable states from the FIB states is given by the endpoints 
    \begin{align*}
        \mathcal{A}[x,y]=\left[0,\frac{\sqrt{3}}{4}\right]\quad\mathrm{and }\quad \mathcal{B}[x,y]=\left[\frac{1}{8},0\right], 
    \end{align*}
    and can be parametrized by $x\in [0,\frac{1}{8}]$ and $y(x) = \frac{\sqrt{3}}{4} (1-8x)$~\cite{EltschkaSiewert2012}. Note that this border itself belongs to the fully separable states, while every FIB state fulfils $y(x) > \frac{\sqrt{3}}{4} (1-8x)$. 

    As seen above, the state $\varrho$ is compressible for $k$ copies, if the term $\frac{\ln(3)}{\ln\left(\frac{3 |p_1-p_2|}{1-p_1-p_2}\right)}$ is positive, ensuring the existence of a finite positive $k$. Noting that $p_1 = 1/8(1+8x+4\sqrt{3} y)$ and $p_2 = 1/8 (1-8x+4\sqrt{3}y)$, this is equivalent to $\ln\left(\frac{6x}{3/4 - \sqrt{3}y}\right) > 0$, which is trivially fulfilled for $y > \frac{\sqrt{3}}{4} (1-8x)$ and therefore for every FIB state.

    \qed
    
    Additionally, one finds that the points lying on the border of the fully separable states with $y(x) = \frac{\sqrt{3}}{4} (1-8x)$ converge to the point $\mathcal{A}[0;\frac{\sqrt{3}}{4}]$ exactly alongside this boundary.

\section{Relation to entanglement distillation}

    The Hadamard map in Eq.~\eqref{eq:HadamardMap} is closely related to entanglement distillation (sometimes called purification) schemes. In the standard distillation setting, the central question is whether, given asymptotically many copies of a noisy entangled state, local operations and classical communication (LOCC) can probabilistically produce a smaller number of states arbitrarily close to a pure target entangled state, such as a Bell pair or GHZ state~\cite{BennettBrassardPopescuSchumacherSmolinWooters1996, MuraoPlenioPopescuVedralKnight1998, DuerAschauerBriegel2003, DuerBriegel2007, HorodeckiRudnickiZyczkowski2022}. Such protocols typically require more general local operations and classical communication to ensure that the relevant target fidelity increases over iterations. However, the question addressed in the present work is conceptually distinct from that of standard distillation schemes and is strictly weaker. Rather than asking whether the repeated application of LOCC asymptotically leads to a specific pure target state, we ask whether the entanglement distributed across multiple copies can be \emph{compressed} back to the single-copy Hilbert space. So, the objective is to ensure that any GME is accessible in the original Hilbert space without a specific target state.

    We note that the Hadamard map was considered in distillation schemes before. In Ref.~\cite{HuberPlesch2011}, the authors consider as a target the GHZ state, and analyse the distillation properties of GHZ states with added white noise. They report that these states are distillable by iteratively applying the Hadamard map, with input fidelities of $0.8$ and $0.5$, achieving arbitrarily high output GHZ fidelity. Our results, on the contrary, show that arbitrarily high fidelity is not achievable in either case. Compressing a state with an input fidelity of $F_{in}=0.8$, one finds that the maximal output GHZ fidelity is achieved for $k=2$ copies, with an output fidelity of $F_{out} \approx 0.92$. Increasing the number of copies decreases output fidelity. For $F_{in} = 0.5$, compression of the state reaches a maximal fidelity of $F_{out} \approx 0.68$, attained at $k=3$, and converges to a GHZ fidelity of $1/2$, i.e. a fully separable state, as $k \to \infty$. However, high target fidelities are attainable by iterating the Hadamard map for other noise models~\cite{huberdiscussion} --- for instance, for the class $\chi(\lambda_1, \lambda_2, \lambda_3, \lambda_4)$ of Sec.~\ref{Sec:GHZDiagonal}, i.e.\ mixtures of different GHZ-type states rather than white noise, the GHZ fidelity does converge to $1$ with growing $k$. 

    On the other hand, traditional GHZ distillation schemes~\cite{MuraoPlenioPopescuVedralKnight1998, DuerAschauerBriegel2003} allow for distillation from the GHZ state with added white noise for a noise parameter $p<2/3$, and a more recent distillation scheme~\cite{RozgonyiSzechenyiKalmanKiss2025} shows the distillability of the GHZ state with added white noise for $p<0.77$, remarkably close to the compressibility bound~(\ref{eq:GHZcompressionBound}). This highlights the physical importance of distinguishing between entanglement distillation and compression. While the Hadamard map can indeed purify certain specially structured state families, its repeated application does not, in general, lead to asymptotic distillation. Conversely, it rather suppresses the compressed GME once the optimal number of copies is exceeded.
   
\section{Beyond ideal states: Application to experimental tomographic data}

\begin{figure*}[!tbp]
  \begin{subfigure}[b]{0.3\textwidth}
    \includegraphics[width=\textwidth]{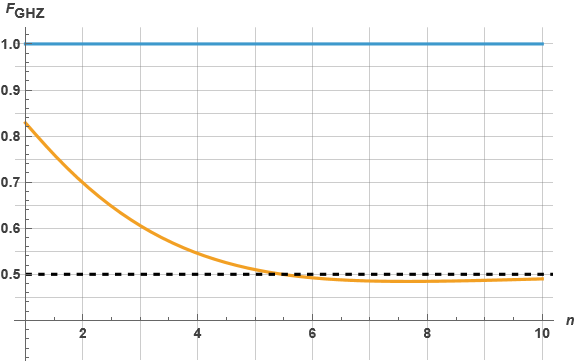}
    \caption{p=0}
  \end{subfigure}
  \hfill
  \begin{subfigure}[b]{0.3\textwidth}
    \includegraphics[width=\textwidth]{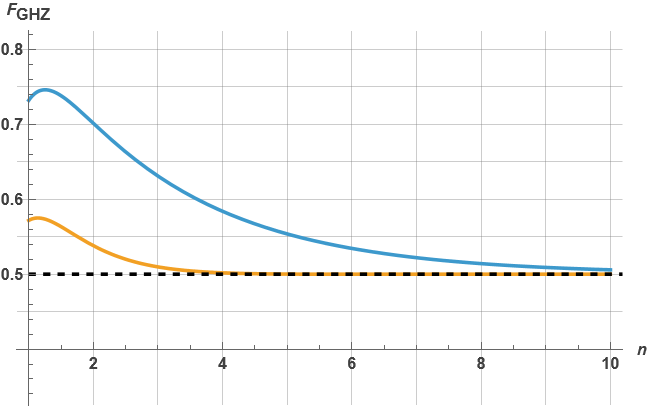}
    \caption{p=0.5}
  \end{subfigure}
  \hfill
  \begin{subfigure}[b]{0.3\textwidth}
    \includegraphics[width=\textwidth]{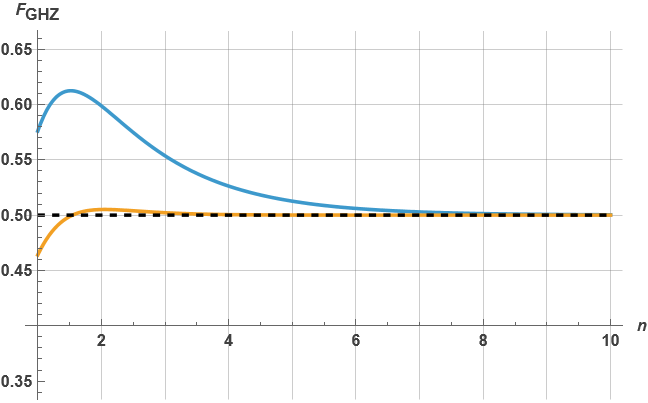}
    \caption{p=0.64}
  \end{subfigure}
  \caption{\justifying GHZ fidelity after iterative Hadamard-map compression of copies drawn from two independent sources, as a function of the number of rounds of $n$. At round $n$, one copy from each of the two sources is compressed via the Hadamard map, so that $n$ rounds correspond to $n$ matched pairs, and $2n$ states in total. The blue curve shows the compression of $2n$ identical copies of the target state, while the orange curve shows the compression of the tomographically reconstructed copies of the two sources from Ref.~\cite{ZhangFeiLiuetal2026}. The dashed line marks $F_{\mathrm{GHZ}} = 1/2$. For case (a), the compression scheme using the realistic data decreases the GHZ fidelity already in the first round, while in case (b), the compression should be stopped after the first round. In case (c), both the theoretical target state and the experimentally prepared states achieve optimal compression after two rounds of the protocol. However, in terms of copies, the theoretical target state reaches the compressibility optimum when three copies are compressed.
  }
  \label{fig:TomographicData}
\end{figure*}

    The states investigated so far satisfy the relevant class conditions exactly, which is an idealisation that cannot be guaranteed in realistic experimental scenarios. To address this, we turn to more physically grounded states by making use of the tomographic data presented in Ref.~\cite{ZhangFeiLiuetal2026}, where two-copy GME superactivation and compression via the Hadamard map were demonstrated experimentally. In this work, two copies of the three-photon state $\varrho(p) = (1-p) \ketbra{\mathrm{GHZ}}{\mathrm{GHZ}} + \frac{p}{8} \mathbb{1}_8$ were prepared using two distinct sources, resulting in two slightly different states. The Hadamard map is then implemented using three polarising beam splitters. To compare the results of this experiment to our analysis above, we consider $n$ rounds of the experiment, which means that $2n$ states are prepared and are pairwise compressed by the Hadamard map. As a quantifier, we again take the GHZ fidelity, yielding the results shown in Fig.~\ref{fig:TomographicData}.

    We find that for $2n$ copies of the ideal state $\varrho(p) = (1-p) \ketbra{\mathrm{GHZ}}{\mathrm{GHZ}} + \frac{p}{8} \mathbb{1}_8$, the highest GHZ fidelity can be reached for $n$ arbitrary ($p=0$), $n=1$ ($p=0.5$) or $n=2$ ($p=0.64$) iterations of the experimental protocol. For the tomographically reconstructed experimental states, we find that for $p=0$, even a single round of the protocol decreases fidelity relative to the initial states; for $p=0.5$, the second round of the protocol already reduces the GHZ fidelity. For the case of $p=0.64$, the state reaches the highest GHZ fidelity after exactly two iterations of the protocol, achieving GHZ fidelity $F_\mathrm{GHZ}=0.505075$. In comparison, in the case of the ideal target state, the highest fidelity is obtained after compressing three copies of the state, achieving GHZ fidelity $F_\mathrm{GHZ}=0.61239$. The difference in achievable fidelities highlights the compression protocol's sensitivity to experimental imperfections.

\section{Conclusion}

    In this study, we examined the compressibility of GME under the Hadamard map in a certain subclass of GHZ-diagonal states and in the GHZ-symmetric states in the many-copy regime. For the subclass of GHZ-diagonal states, we found that they are compressible for some finite number of copies, except for a subset that has measure zero. We found that all GME-activatable GHZ-symmetric states are also compressible under the Hadamard map when sufficiently many copies are used. Further, we found that it is not only important to define the minimal number of copies but also to estimate the maximal number of copies required to compress entanglement via the Hadamard map, as the fidelity with the GHZ state begins to decrease beyond an optimal number of copies. It always converges to 1/2 as the states converge to fully separable states in the limit of infinitely many copies. Therefore, for practical use, one needs to either determine the optimal number of copies or use another strategy, depending on the task at hand.

\section*{Acknowledgments}
We thank Sophia Denker, Nicolai Friis, Otfried G\"uhne, Marcus Huber, and Lina Vandr\'e for fruitful discussions.
K.B. acknowledges financial support from the Austrian Science Fund (FWF) through the project P 36478-N funded by the European Union{\textemdash}NextGenerationEU, and from the Czech Science Foundation through the Junior Star grant 25-17250M.
L.T.W. acknowledges financial support from the Deutsche Forschungsgemeinschaft (DFG, German Research Foundation, project number 563437167), the Sino-German Center for Research Promotion (Project M-0294), and the German Federal Ministry of Research, Technology and Space (Project QuKuK, Grant No. 16KIS1618K and Project BeRyQC, Grant No. 13N17292).
L.T.W. also acknowledges support from the House of Young Talents of the University of Siegen.

\appendix

\section{Stability analysis with Euler method}\label{app-sec-EulerMethod}

    We want to demonstrate the relation between the fixed points of a vector field and those of the map approximating it by the Euler method (see, e.g.,~\cite [Chapter 2.8]{Strogatz1994}).

    Let us have a discrete map that transforms the coordinates $[x,y]$ to updated coordinates $[x',y']$ as
    \begin{align*}
        \mathbf{X'}=\begin{pmatrix} x' \\ y' \end{pmatrix}=\begin{pmatrix}
            f_1(x,y) \\ f_2(x,y)
        \end{pmatrix}=\mathbf{f}\left(\mathbf{X}\right).
    \end{align*}
    Then, if we have a system of ODE $\dot{\mathbf{X}}=\mathbf{F}(\mathbf{X})$, its Euler's approximation with a step $\Delta t$ is
    \begin{align*}
        \mathbf{X}_{n+1}=\mathbf{X}_n+\Delta t\mathbf{F}(\mathbf{X}_n),
    \end{align*}
    so
    \begin{align*}
        \mathbf{f}(\mathbf{X})&=\mathbf{X}+\Delta t\mathbf{F}(\mathbf{X}) \\
        \mathbf{F}(\mathbf{X})&=\frac{\mathbf{f}(\mathbf{X})-\mathbf{X}}{\Delta t}=\dot{\mathbf{X}},
    \end{align*}
    and therefore for the fixed points we have
    \begin{align*}
        \dot{\mathbf{X}}=0\quad\Rightarrow\quad\mathbf{f}(\mathbf{X})=\mathbf{X}.
    \end{align*}

    Moreover, for the maps considered here, the stability classification of these fixed points is preserved under this correspondence.


\bibliographystyle{apsrev4-1fixed_with_article_titles_full_names_new}
\bibliography{references}

\begin{thebibliography}{30}%
\makeatletter
\providecommand \@ifxundefined [1]{%
 \@ifx{#1\undefined}
}%
\providecommand \@ifnum [1]{%
 \ifnum #1\expandafter \@firstoftwo
 \else \expandafter \@secondoftwo
 \fi
}%
\providecommand \@ifx [1]{%
 \ifx #1\expandafter \@firstoftwo
 \else \expandafter \@secondoftwo
 \fi
}%
\providecommand \natexlab [1]{#1}%
\providecommand \enquote  [1]{#1}%
\providecommand \bibnamefont  [1]{#1}%
\providecommand \bibfnamefont [1]{#1}%
\providecommand \citenamefont [1]{#1}%
\providecommand \href@noop [0]{\@secondoftwo}%
\providecommand \href [0]{\begingroup \@sanitize@url \@href}%
\providecommand \@href[1]{\@@startlink{#1}\@@href}%
\providecommand \@@href[1]{\endgroup#1\@@endlink}%
\providecommand \@sanitize@url [0]{\catcode `\\12\catcode `\$12\catcode `\&12\catcode `\#12\catcode `\^12\catcode `\_12\catcode `\%12\relax}%
\providecommand \@@startlink[1]{}%
\providecommand \@@endlink[0]{}%
\providecommand \url  [0]{\begingroup\@sanitize@url \@url }%
\providecommand \@url [1]{\endgroup\@href {#1}{\urlprefix }}%
\providecommand \urlprefix  [0]{URL }%
\providecommand \Eprint [0]{\href }%
\providecommand \doibase [0]{https://doi.org/}%
\providecommand \selectlanguage [0]{\@gobble}%
\providecommand \bibinfo  [0]{\@secondoftwo}%
\providecommand \bibfield  [0]{\@secondoftwo}%
\providecommand \translation [1]{[#1]}%
\providecommand \BibitemOpen [0]{}%
\providecommand \bibitemStop [0]{}%
\providecommand \bibitemNoStop [0]{.\EOS\space}%
\providecommand \EOS [0]{\spacefactor3000\relax}%
\providecommand \BibitemShut  [1]{\csname bibitem#1\endcsname}%
\let\auto@bib@innerbib\@empty
\bibitem [{\citenamefont {T\'oth}(2012)}]{Toth2012}%
  \BibitemOpen
  \bibfield  {author} {\bibinfo {author} {\bibfnamefont {G\'eza}\ \bibnamefont {T\'oth}},\ }\emph {\enquote {\bibinfo {title} {Multipartite entanglement and high-precision metrology},}\ }\href {\doibase 10.1103/PhysRevA.85.022322} {\bibfield  {journal} {\bibinfo  {journal} {Phys. Rev. A}\ }\textbf {\bibinfo {volume} {85}},\ \bibinfo {pages} {022322} (\bibinfo {year} {2012})},\ \Eprint {http://arxiv.org/abs/1006.4368} {arXiv:1006.4368}\BibitemShut {NoStop}%
\bibitem [{\citenamefont {Hyllus}\ \emph {et~al.}(2012)\citenamefont {Hyllus}, \citenamefont {Laskowski}, \citenamefont {Krischek}, \citenamefont {Schwemmer}, \citenamefont {Wieczorek}, \citenamefont {Weinfurter}, \citenamefont {Pezz{\`e}},\ and\ \citenamefont {Smerzi}}]{HyllusLaskowskiKrischekEtal2012}%
  \BibitemOpen
  \bibfield  {author} {\bibinfo {author} {\bibfnamefont {Philipp}\ \bibnamefont {Hyllus}}, \bibinfo {author} {\bibfnamefont {Wies{\l}aw}\ \bibnamefont {Laskowski}}, \bibinfo {author} {\bibfnamefont {Roland}\ \bibnamefont {Krischek}}, \bibinfo {author} {\bibfnamefont {Christian}\ \bibnamefont {Schwemmer}}, \bibinfo {author} {\bibfnamefont {Witlef}\ \bibnamefont {Wieczorek}}, \bibinfo {author} {\bibfnamefont {Harald}\ \bibnamefont {Weinfurter}}, \bibinfo {author} {\bibfnamefont {Luca}\ \bibnamefont {Pezz{\`e}}}, \ and\ \bibinfo {author} {\bibfnamefont {Augusto}\ \bibnamefont {Smerzi}},\ }\emph {\enquote {\bibinfo {title} {Fisher information and multiparticle entanglement},}\ }\href {https://doi.org/10.1103/PhysRevA.85.022321} {\bibfield  {journal} {\bibinfo  {journal} {Phys. Rev. A}\ }\textbf {\bibinfo {volume} {85}},\ \bibinfo {pages} {022321} (\bibinfo {year} {2012})},\ \Eprint {http://arxiv.org/abs/1006.4366} {arXiv:1006.4366}\BibitemShut {NoStop}%
\bibitem [{\citenamefont {Hillery}\ \emph {et~al.}(1999)\citenamefont {Hillery}, \citenamefont {Bu\ifmmode~\check{z}\else \v{z}\fi{}ek},\ and\ \citenamefont {Berthiaume}}]{HilleryBuzekBerthiaume1999}%
  \BibitemOpen
  \bibfield  {author} {\bibinfo {author} {\bibfnamefont {Mark}\ \bibnamefont {Hillery}}, \bibinfo {author} {\bibfnamefont {Vladim\'{\i}r}\ \bibnamefont {Bu\ifmmode~\check{z}\else \v{z}\fi{}ek}}, \ and\ \bibinfo {author} {\bibfnamefont {Andr\'e}\ \bibnamefont {Berthiaume}},\ }\emph {\enquote {\bibinfo {title} {Quantum secret sharing},}\ }\href {\doibase 10.1103/PhysRevA.59.1829} {\bibfield  {journal} {\bibinfo  {journal} {Phys. Rev. A}\ }\textbf {\bibinfo {volume} {59}},\ \bibinfo {pages} {1829} (\bibinfo {year} {1999})},\ \Eprint {http://arxiv.org/abs/quant-ph/9806063} {arXiv:quant-ph/9806063}\BibitemShut {NoStop}%
\bibitem [{\citenamefont {Schauer}\ \emph {et~al.}(2010)\citenamefont {Schauer}, \citenamefont {Huber},\ and\ \citenamefont {Hiesmayr}}]{SchauerHuberHiesmayr2010}%
  \BibitemOpen
  \bibfield  {author} {\bibinfo {author} {\bibfnamefont {Stefan}\ \bibnamefont {Schauer}}, \bibinfo {author} {\bibfnamefont {Marcus}\ \bibnamefont {Huber}}, \ and\ \bibinfo {author} {\bibfnamefont {Beatrix~C.}\ \bibnamefont {Hiesmayr}},\ }\emph {\enquote {\bibinfo {title} {Experimentally feasible security check for $n$-qubit quantum secret sharing},}\ }\href {\doibase 10.1103/PhysRevA.82.062311} {\bibfield  {journal} {\bibinfo  {journal} {Phys. Rev. A}\ }\textbf {\bibinfo {volume} {82}},\ \bibinfo {pages} {062311} (\bibinfo {year} {2010})},\ \Eprint {http://arxiv.org/abs/1009.4796} {arXiv:1009.4796}\BibitemShut {NoStop}%
\bibitem [{\citenamefont {Cao}\ \emph {et~al.}(2023)\citenamefont {Cao}, \citenamefont {Wu}, \citenamefont {Chen}, \citenamefont {Gong}, \citenamefont {Wu}, \citenamefont {Ye}, \citenamefont {Zha}, \citenamefont {Qian}, \citenamefont {Ying}, \citenamefont {Guo}, \citenamefont {Zhu}, \citenamefont {Huang}, \citenamefont {Zhao}, \citenamefont {Li}, \citenamefont {Wang}, \citenamefont {Yu}, \citenamefont {Fan}, \citenamefont {Wu}, \citenamefont {Su}, \citenamefont {Deng}, \citenamefont {Rong}, \citenamefont {Li}, \citenamefont {Zhang}, \citenamefont {Chung}, \citenamefont {Liang}, \citenamefont {Lin}, \citenamefont {Xu}, \citenamefont {Sun}, \citenamefont {Guo}, \citenamefont {Li}, \citenamefont {Huo}, \citenamefont {Peng}, \citenamefont {Lu}, \citenamefont {Yuan}, \citenamefont {Zhu},\ and\ \citenamefont {Pan}}]{CaoEtAl2023}%
  \BibitemOpen
  \bibfield  {author} {\bibinfo {author} {\bibfnamefont {Sirui}\ \bibnamefont {Cao}}, \bibinfo {author} {\bibfnamefont {Bujiao}\ \bibnamefont {Wu}}, \bibinfo {author} {\bibfnamefont {Fusheng}\ \bibnamefont {Chen}}, \bibinfo {author} {\bibfnamefont {Ming}\ \bibnamefont {Gong}}, \bibinfo {author} {\bibfnamefont {Yulin}\ \bibnamefont {Wu}}, \bibinfo {author} {\bibfnamefont {Yangsen}\ \bibnamefont {Ye}}, \bibinfo {author} {\bibfnamefont {Chen}\ \bibnamefont {Zha}}, \bibinfo {author} {\bibfnamefont {Haoran}\ \bibnamefont {Qian}}, \bibinfo {author} {\bibfnamefont {Chong}\ \bibnamefont {Ying}}, \bibinfo {author} {\bibfnamefont {Shaojun}\ \bibnamefont {Guo}}, \bibinfo {author} {\bibfnamefont {Qingling}\ \bibnamefont {Zhu}}, \bibinfo {author} {\bibfnamefont {He-Liang}\ \bibnamefont {Huang}}, \bibinfo {author} {\bibfnamefont {Youwei}\ \bibnamefont {Zhao}}, \bibinfo {author} {\bibfnamefont {Shaowei}\ \bibnamefont {Li}}, \bibinfo {author} {\bibfnamefont {Shiyu}\ \bibnamefont {Wang}}, \bibinfo {author} {\bibfnamefont
  {Jiale}\ \bibnamefont {Yu}}, \bibinfo {author} {\bibfnamefont {Daojin}\ \bibnamefont {Fan}}, \bibinfo {author} {\bibfnamefont {Dachao}\ \bibnamefont {Wu}}, \bibinfo {author} {\bibfnamefont {Hong}\ \bibnamefont {Su}}, \bibinfo {author} {\bibfnamefont {Hui}\ \bibnamefont {Deng}}, \bibinfo {author} {\bibfnamefont {Hao}\ \bibnamefont {Rong}}, \bibinfo {author} {\bibfnamefont {Yuan}\ \bibnamefont {Li}}, \bibinfo {author} {\bibfnamefont {Kaili}\ \bibnamefont {Zhang}}, \bibinfo {author} {\bibfnamefont {Tung-Hsun}\ \bibnamefont {Chung}}, \bibinfo {author} {\bibfnamefont {Futian}\ \bibnamefont {Liang}}, \bibinfo {author} {\bibfnamefont {Jin}\ \bibnamefont {Lin}}, \bibinfo {author} {\bibfnamefont {Yu}~\bibnamefont {Xu}}, \bibinfo {author} {\bibfnamefont {Lihua}\ \bibnamefont {Sun}}, \bibinfo {author} {\bibfnamefont {Cheng}\ \bibnamefont {Guo}}, \bibinfo {author} {\bibfnamefont {Na}~\bibnamefont {Li}}, \bibinfo {author} {\bibfnamefont {Yong-Heng}\ \bibnamefont {Huo}}, \bibinfo {author} {\bibfnamefont {Cheng-Zhi}\
  \bibnamefont {Peng}}, \bibinfo {author} {\bibfnamefont {Chao-Yang}\ \bibnamefont {Lu}}, \bibinfo {author} {\bibfnamefont {Xiao}\ \bibnamefont {Yuan}}, \bibinfo {author} {\bibfnamefont {Xiaobo}\ \bibnamefont {Zhu}}, \ and\ \bibinfo {author} {\bibfnamefont {Jian-Wei}\ \bibnamefont {Pan}},\ }\emph {\enquote {\bibinfo {title} {{Generation of genuine entanglement up to 51 superconducting qubits}},}\ }\href {https://doi.org/10.1038/s41586-023-06195-1} {\bibfield  {journal} {\bibinfo  {journal} {Nature}\ }\textbf {\bibinfo {volume} {619}},\ \bibinfo {pages} {738{\textendash}742} (\bibinfo {year} {2023})}\BibitemShut {NoStop}%
\bibitem [{\citenamefont {Yamasaki}\ \emph {et~al.}(2022)\citenamefont {Yamasaki}, \citenamefont {Morelli}, \citenamefont {Miethlinger}, \citenamefont {Bavaresco}, \citenamefont {Friis},\ and\ \citenamefont {Huber}}]{YamasakiMorelliMiethlingerBavarescoFriisHuber2022}%
  \BibitemOpen
  \bibfield  {author} {\bibinfo {author} {\bibfnamefont {Hayata}\ \bibnamefont {Yamasaki}}, \bibinfo {author} {\bibfnamefont {Simon}\ \bibnamefont {Morelli}}, \bibinfo {author} {\bibfnamefont {Markus}\ \bibnamefont {Miethlinger}}, \bibinfo {author} {\bibfnamefont {Jessica}\ \bibnamefont {Bavaresco}}, \bibinfo {author} {\bibfnamefont {Nicolai}\ \bibnamefont {Friis}}, \ and\ \bibinfo {author} {\bibfnamefont {Marcus}\ \bibnamefont {Huber}},\ }\emph {\enquote {\bibinfo {title} {{Activation of genuine multipartite entanglement: beyond the single-copy paradigm of entanglement characterisation}},}\ }\href {https://doi.org/10.22331/q-2022-04-25-695} {\bibfield  {journal} {\bibinfo  {journal} {Quantum}\ }\textbf {\bibinfo {volume} {6}},\ \bibinfo {pages} {695} (\bibinfo {year} {2022})},\ \Eprint {http://arxiv.org/abs/2106.01372} {arXiv:2106.01372}\BibitemShut {NoStop}%
\bibitem [{\citenamefont {Palazuelos}\ and\ \citenamefont {de~Vicente}(2022)}]{PalazuelosDeVicente2022}%
  \BibitemOpen
  \bibfield  {author} {\bibinfo {author} {\bibfnamefont {Carlos}\ \bibnamefont {Palazuelos}}\ and\ \bibinfo {author} {\bibfnamefont {Julio~I.}\ \bibnamefont {de~Vicente}},\ }\emph {\enquote {\bibinfo {title} {{Genuine multipartite entanglement of quantum states in the multiple-copy scenario}},}\ }\href {https://doi.org/10.22331/q-2022-06-13-735} {\bibfield  {journal} {\bibinfo  {journal} {Quantum}\ }\textbf {\bibinfo {volume} {6}},\ \bibinfo {pages} {735} (\bibinfo {year} {2022})},\ \Eprint {http://arxiv.org/abs/2201.08694} {arXiv:2201.08694}\BibitemShut {NoStop}%
\bibitem [{\citenamefont {Baksov{\'{a}}}\ \emph {et~al.}(2025)\citenamefont {Baksov{\'{a}}}, \citenamefont {Leskovjanov{\'{a}}}, \citenamefont {Mi{\v{s}}ta~Jr.}, \citenamefont {Agudelo},\ and\ \citenamefont {Friis}}]{BaksovaLeskovjanovaMistaAgudeloFriis2024}%
  \BibitemOpen
  \bibfield  {author} {\bibinfo {author} {\bibfnamefont {Kl{\'{a}}ra}\ \bibnamefont {Baksov{\'{a}}}}, \bibinfo {author} {\bibfnamefont {Olga}\ \bibnamefont {Leskovjanov{\'{a}}}}, \bibinfo {author} {\bibfnamefont {Ladislav}\ \bibnamefont {Mi{\v{s}}ta~Jr.}}, \bibinfo {author} {\bibfnamefont {Elizabeth}\ \bibnamefont {Agudelo}}, \ and\ \bibinfo {author} {\bibfnamefont {Nicolai}\ \bibnamefont {Friis}},\ }\emph {\enquote {\bibinfo {title} {Multi-copy activation of genuine multipartite entanglement in continuous-variable systems},}\ }\href {\doibase 10.22331/q-2025-04-09-1699} {\bibfield  {journal} {\bibinfo  {journal} {{Quantum}}\ }\textbf {\bibinfo {volume} {9}},\ \bibinfo {pages} {1699} (\bibinfo {year} {2025})},\ \Eprint {http://arxiv.org/abs/2312.16570} {arXiv:2312.16570}\BibitemShut {NoStop}%
\bibitem [{\citenamefont {Zhang}\ \emph {et~al.}(2026)\citenamefont {Zhang}, \citenamefont {Fei}, \citenamefont {Liu}, \citenamefont {Zhang}, \citenamefont {Yin}, \citenamefont {Mao}, \citenamefont {Li}, \citenamefont {Liu}, \citenamefont {G\"uhne}, \citenamefont {Ma}, \citenamefont {Chen},\ and\ \citenamefont {Pan}}]{ZhangFeiLiuetal2026}%
  \BibitemOpen
  \bibfield  {author} {\bibinfo {author} {\bibfnamefont {Rui}\ \bibnamefont {Zhang}}, \bibinfo {author} {\bibfnamefont {Yue-Yang}\ \bibnamefont {Fei}}, \bibinfo {author} {\bibfnamefont {Zhenhuan}\ \bibnamefont {Liu}}, \bibinfo {author} {\bibfnamefont {Xingjian}\ \bibnamefont {Zhang}}, \bibinfo {author} {\bibfnamefont {Xu-Fei}\ \bibnamefont {Yin}}, \bibinfo {author} {\bibfnamefont {Yingqiu}\ \bibnamefont {Mao}}, \bibinfo {author} {\bibfnamefont {Li}~\bibnamefont {Li}}, \bibinfo {author} {\bibfnamefont {Nai-Le}\ \bibnamefont {Liu}}, \bibinfo {author} {\bibfnamefont {Otfried}\ \bibnamefont {G\"uhne}}, \bibinfo {author} {\bibfnamefont {Xiongfeng}\ \bibnamefont {Ma}}, \bibinfo {author} {\bibfnamefont {Yu-Ao}\ \bibnamefont {Chen}}, \ and\ \bibinfo {author} {\bibfnamefont {Jian-Wei}\ \bibnamefont {Pan}},\ }\emph {\enquote {\bibinfo {title} {Entanglement superactivation in multiphoton distillation networks},}\ }\href {\doibase 10.1103/zkd3-4gmx} {\bibfield  {journal} {\bibinfo  {journal} {Phys. Rev. Lett.}\ }\textbf
  {\bibinfo {volume} {136}},\ \bibinfo {pages} {120801} (\bibinfo {year} {2026})},\ \Eprint {http://arxiv.org/abs/2510.26290} {arXiv:2510.26290}\BibitemShut {NoStop}%
\bibitem [{\citenamefont {St\'{a}rek}\ \emph {et~al.}(2026)\citenamefont {St\'{a}rek}, \citenamefont {Gollerthan}, \citenamefont {Leskovjanov{\'a}}, \citenamefont {Meth}, \citenamefont {Tirler}, \citenamefont {Friis}, \citenamefont {Ringbauer},\ and\ \citenamefont {Mi\v{s}ta~Jr.}}]{StarekGollerthanLeskovjanovaMethTirlerFriisRingbauerMista2026}%
  \BibitemOpen
  \bibfield  {author} {\bibinfo {author} {\bibfnamefont {Robert}\ \bibnamefont {St\'{a}rek}}, \bibinfo {author} {\bibfnamefont {Tim}\ \bibnamefont {Gollerthan}}, \bibinfo {author} {\bibfnamefont {Olga}\ \bibnamefont {Leskovjanov{\'a}}}, \bibinfo {author} {\bibfnamefont {Michael}\ \bibnamefont {Meth}}, \bibinfo {author} {\bibfnamefont {Peter}\ \bibnamefont {Tirler}}, \bibinfo {author} {\bibfnamefont {Nicolai}\ \bibnamefont {Friis}}, \bibinfo {author} {\bibfnamefont {Martin}\ \bibnamefont {Ringbauer}}, \ and\ \bibinfo {author} {\bibfnamefont {Ladislav}\ \bibnamefont {Mi\v{s}ta~Jr.}},\ }\emph {\enquote {\bibinfo {title} {{Experimental Verification of Multicopy Activation of Genuine Multipartite Entanglement}},}\ }\href {https://doi.org/10.1103/kv4s-tfc6} {\bibfield  {journal} {\bibinfo  {journal} {Phys. Rev. Lett.}\ }\textbf {\bibinfo {volume} {136}},\ \bibinfo {pages} {160201} (\bibinfo {year} {2026})},\ \Eprint {http://arxiv.org/abs/2510.12457} {arXiv:2510.12457}\BibitemShut {NoStop}%
\bibitem [{\citenamefont {Weinbrenner}\ \emph {et~al.}(2024)\citenamefont {Weinbrenner}, \citenamefont {Baksov\'a}, \citenamefont {Denker}, \citenamefont {Morelli}, \citenamefont {Yu}, \citenamefont {Friis},\ and\ \citenamefont {G\"uhne}}]{WeinbrennerBaksovaDenkerMorelliYuFriisGuhne2024}%
  \BibitemOpen
  \bibfield  {author} {\bibinfo {author} {\bibfnamefont {Lisa~T.}\ \bibnamefont {Weinbrenner}}, \bibinfo {author} {\bibfnamefont {Kl\'ara}\ \bibnamefont {Baksov\'a}}, \bibinfo {author} {\bibfnamefont {Sophia}\ \bibnamefont {Denker}}, \bibinfo {author} {\bibfnamefont {Simon}\ \bibnamefont {Morelli}}, \bibinfo {author} {\bibfnamefont {Xiao-Dong}\ \bibnamefont {Yu}}, \bibinfo {author} {\bibfnamefont {Nicolai}\ \bibnamefont {Friis}}, \ and\ \bibinfo {author} {\bibfnamefont {Otfried}\ \bibnamefont {G\"uhne}},\ }\href@noop {} {\emph {\enquote {\bibinfo {title} {Superactivation and incompressibility of genuine multipartite entanglement},}\ }}\Eprint {http://arxiv.org/abs/2412.18331} {arXiv:2412.18331} [quant-ph] (\bibinfo {year} {2024})\BibitemShut {NoStop}%
\bibitem [{\citenamefont {G{\"u}hne}\ and\ \citenamefont {Seevinck}(2010)}]{GuehneSeevinck2010}%
  \BibitemOpen
  \bibfield  {author} {\bibinfo {author} {\bibfnamefont {Otfried}\ \bibnamefont {G{\"u}hne}}\ and\ \bibinfo {author} {\bibfnamefont {Michael}\ \bibnamefont {Seevinck}},\ }\emph {\enquote {\bibinfo {title} {Separability criteria for genuine multiparticle entanglement},}\ }\href {https://doi.org/10.1088/1367-2630/12/5/053002} {\bibfield  {journal} {\bibinfo  {journal} {New J. Phys.}\ }\textbf {\bibinfo {volume} {12}},\ \bibinfo {pages} {053002} (\bibinfo {year} {2010})},\ \Eprint {http://arxiv.org/abs/0905.1349} {arXiv:0905.1349}\BibitemShut {NoStop}%
\bibitem [{\citenamefont {D{\"u}r}\ and\ \citenamefont {Cirac}(2001)}]{DuerCirac2001}%
  \BibitemOpen
  \bibfield  {author} {\bibinfo {author} {\bibfnamefont {W.}~\bibnamefont {D{\"u}r}}\ and\ \bibinfo {author} {\bibfnamefont {J.~I.}\ \bibnamefont {Cirac}},\ }\emph {\enquote {\bibinfo {title} {Multiparticle entanglement and its experimental detection},}\ }\href {https://doi.org/10.1088/0305-4470/34/35/310} {\bibfield  {journal} {\bibinfo  {journal} {J. Phys. A: Math. Gen.}\ }\textbf {\bibinfo {volume} {34}},\ \bibinfo {pages} {6837} (\bibinfo {year} {2001})},\ \Eprint {http://arxiv.org/abs/quant-ph/0011025} {arXiv:quant-ph/0011025}\BibitemShut {NoStop}%
\bibitem [{\citenamefont {Eltschka}\ and\ \citenamefont {Siewert}(2012{\natexlab{a}})}]{EltschkaSiewert2012}%
  \BibitemOpen
  \bibfield  {author} {\bibinfo {author} {\bibfnamefont {Christopher}\ \bibnamefont {Eltschka}}\ and\ \bibinfo {author} {\bibfnamefont {Jens}\ \bibnamefont {Siewert}},\ }\emph {\enquote {\bibinfo {title} {Entanglement of three-qubit greenberger-horne-zeilinger--symmetric states},}\ }\href {\doibase 10.1103/PhysRevLett.108.020502} {\bibfield  {journal} {\bibinfo  {journal} {Phys. Rev. Lett.}\ }\textbf {\bibinfo {volume} {108}},\ \bibinfo {pages} {020502} (\bibinfo {year} {2012}{\natexlab{a}})},\ \Eprint {http://arxiv.org/abs/1304.6095} {arXiv:1304.6095}\BibitemShut {NoStop}%
\bibitem [{\citenamefont {Eltschka}\ and\ \citenamefont {Siewert}(2012{\natexlab{b}})}]{EltschkaSiewert_2012b}%
  \BibitemOpen
  \bibfield  {author} {\bibinfo {author} {\bibfnamefont {Christopher}\ \bibnamefont {Eltschka}}\ and\ \bibinfo {author} {\bibfnamefont {Jens}\ \bibnamefont {Siewert}},\ }\emph {\enquote {\bibinfo {title} {{A quantitative witness for Greenberger-Horne-Zeilinger entanglement}},}\ }\href {https://doi.org/10.1038/srep00942} {\bibfield  {journal} {\bibinfo  {journal} {Sci. Rep.}\ }\textbf {\bibinfo {volume} {2}},\ \bibinfo {pages} {942} (\bibinfo {year} {2012}{\natexlab{b}})},\ \Eprint {http://arxiv.org/abs/1305.1930} {arXiv:1305.1930}\BibitemShut {NoStop}%
\bibitem [{\citenamefont {Eltschka}\ and\ \citenamefont {Siewert}(2013)}]{EltschkaSiewert_2012c}%
  \BibitemOpen
  \bibfield  {author} {\bibinfo {author} {\bibfnamefont {Christopher}\ \bibnamefont {Eltschka}}\ and\ \bibinfo {author} {\bibfnamefont {Jens}\ \bibnamefont {Siewert}},\ }\emph {\enquote {\bibinfo {title} {{Optimal witnesses for three-qubit entanglement from Greenberger-Horne-Zeilinger symmetry}},}\ }\href {https://doi.org/10.26421/QIC13.3-4-3} {\bibfield  {journal} {\bibinfo  {journal} {Quant. Inf. Comput.}\ }\textbf {\bibinfo {volume} {13}},\ \bibinfo {pages} {210} (\bibinfo {year} {2013})},\ \Eprint {http://arxiv.org/abs/1204.5451} {arXiv:1204.5451}\BibitemShut {NoStop}%
\bibitem [{\citenamefont {G{\"u}hne}\ and\ \citenamefont {T{\'o}th}(2009)}]{GuehneToth2009}%
  \BibitemOpen
  \bibfield  {author} {\bibinfo {author} {\bibfnamefont {Otfried}\ \bibnamefont {G{\"u}hne}}\ and\ \bibinfo {author} {\bibfnamefont {G{\'e}za}\ \bibnamefont {T{\'o}th}},\ }\emph {\enquote {\bibinfo {title} {Entanglement detection},}\ }\href {https://doi.org/10.1016/j.physrep.2009.02.004} {\bibfield  {journal} {\bibinfo  {journal} {Phys. Rep.}\ }\textbf {\bibinfo {volume} {474}},\ \bibinfo {pages} {1} (\bibinfo {year} {2009})},\ \Eprint {http://arxiv.org/abs/0811.2803} {arXiv:0811.2803}\BibitemShut {NoStop}%
\bibitem [{\citenamefont {Friis}\ \emph {et~al.}(2019)\citenamefont {Friis}, \citenamefont {Vitagliano}, \citenamefont {Malik},\ and\ \citenamefont {Huber}}]{FriisVitaglianoMalikHuber2019}%
  \BibitemOpen
  \bibfield  {author} {\bibinfo {author} {\bibfnamefont {Nicolai}\ \bibnamefont {Friis}}, \bibinfo {author} {\bibfnamefont {Giuseppe}\ \bibnamefont {Vitagliano}}, \bibinfo {author} {\bibfnamefont {Mehul}\ \bibnamefont {Malik}}, \ and\ \bibinfo {author} {\bibfnamefont {Marcus}\ \bibnamefont {Huber}},\ }\emph {\enquote {\bibinfo {title} {{Entanglement Certification From Theory to Experiment}},}\ }\href {https://doi.org/10.1038/s42254-018-0003-5} {\bibfield  {journal} {\bibinfo  {journal} {Nat. Rev. Phys.}\ }\textbf {\bibinfo {volume} {1}},\ \bibinfo {pages} {72} (\bibinfo {year} {2019})},\ \Eprint {http://arxiv.org/abs/1906.10929} {arXiv:1906.10929}\BibitemShut {NoStop}%
\bibitem [{\citenamefont {Huber}\ and\ \citenamefont {Plesch}(2011)}]{HuberPlesch2011}%
  \BibitemOpen
  \bibfield  {author} {\bibinfo {author} {\bibfnamefont {Marcus}\ \bibnamefont {Huber}}\ and\ \bibinfo {author} {\bibfnamefont {Martin}\ \bibnamefont {Plesch}},\ }\emph {\enquote {\bibinfo {title} {Purification of genuine multipartite entanglement},}\ }\href {https://doi.org/10.1103/PhysRevA.83.062321} {\bibfield  {journal} {\bibinfo  {journal} {Phys. Rev. A}\ }\textbf {\bibinfo {volume} {83}},\ \bibinfo {pages} {062321} (\bibinfo {year} {2011})},\ \Eprint {http://arxiv.org/abs/1103.4294} {arXiv:1103.4294}\BibitemShut {NoStop}%
\bibitem [{\citenamefont {Lami}\ and\ \citenamefont {Huber}(2016)}]{LamiHuber2016}%
  \BibitemOpen
  \bibfield  {author} {\bibinfo {author} {\bibfnamefont {Ludovico}\ \bibnamefont {Lami}}\ and\ \bibinfo {author} {\bibfnamefont {Marcus}\ \bibnamefont {Huber}},\ }\emph {\enquote {\bibinfo {title} {Bipartite depolarizing channels},}\ }\href {https://doi.org/10.1063/1.4962339} {\bibfield  {journal} {\bibinfo  {journal} {J. Math. Phys.}\ }\textbf {\bibinfo {volume} {57}},\ \bibinfo {pages} {092201} (\bibinfo {year} {2016})},\ \Eprint {http://arxiv.org/abs/1603.02158} {arXiv:1603.02158}\BibitemShut {NoStop}%
\bibitem [{\citenamefont {Holmes}\ \emph {et~al.}(2023)\citenamefont {Holmes}, \citenamefont {Coble}, \citenamefont {Sornborger},\ and\ \citenamefont {Suba{\c{s}}{\i}}}]{HolmesCobleSornborgerSuba2023}%
  \BibitemOpen
  \bibfield  {author} {\bibinfo {author} {\bibfnamefont {Zo\"e}\ \bibnamefont {Holmes}}, \bibinfo {author} {\bibfnamefont {Nolan~J.}\ \bibnamefont {Coble}}, \bibinfo {author} {\bibfnamefont {Andrew~T.}\ \bibnamefont {Sornborger}}, \ and\ \bibinfo {author} {\bibfnamefont {Yi{\u{g}}it}\ \bibnamefont {Suba{\c{s}}{\i}}},\ }\emph {\enquote {\bibinfo {title} {Nonlinear transformations in quantum computation},}\ }\href {\doibase 10.1103/PhysRevResearch.5.013105} {\bibfield  {journal} {\bibinfo  {journal} {Phys. Rev. Res.}\ }\textbf {\bibinfo {volume} {5}},\ \bibinfo {pages} {013105} (\bibinfo {year} {2023})},\ \Eprint {http://arxiv.org/abs/2112.12307} {arXiv:2112.12307}\BibitemShut {NoStop}%
\bibitem [{\citenamefont {Ecker}\ \emph {et~al.}(2021)\citenamefont {Ecker}, \citenamefont {Sohr}, \citenamefont {Bulla}, \citenamefont {Huber}, \citenamefont {Bohmann},\ and\ \citenamefont {Ursin}}]{EckerSohrBullaHuberBohmannUrsin2021}%
  \BibitemOpen
  \bibfield  {author} {\bibinfo {author} {\bibfnamefont {Sebastian}\ \bibnamefont {Ecker}}, \bibinfo {author} {\bibfnamefont {Philipp}\ \bibnamefont {Sohr}}, \bibinfo {author} {\bibfnamefont {Lukas}\ \bibnamefont {Bulla}}, \bibinfo {author} {\bibfnamefont {Marcus}\ \bibnamefont {Huber}}, \bibinfo {author} {\bibfnamefont {Martin}\ \bibnamefont {Bohmann}}, \ and\ \bibinfo {author} {\bibfnamefont {Rupert}\ \bibnamefont {Ursin}},\ }\emph {\enquote {\bibinfo {title} {Experimental single-copy entanglement distillation},}\ }\href {\doibase 10.1103/PhysRevLett.127.040506} {\bibfield  {journal} {\bibinfo  {journal} {Phys. Rev. Lett.}\ }\textbf {\bibinfo {volume} {127}},\ \bibinfo {pages} {040506} (\bibinfo {year} {2021})},\ \Eprint {http://arxiv.org/abs/2101.11503} {arXiv:2101.11503}\BibitemShut {NoStop}%
\bibitem [{\citenamefont {Strogatz}(1994)}]{Strogatz1994}%
  \BibitemOpen
  \bibfield  {author} {\bibinfo {author} {\bibfnamefont {Steven~H.}\ \bibnamefont {Strogatz}},\ }\href@noop {} {\emph {\bibinfo {title} {{Nonlinear Dynamics and Chaos With Applications to Physics, Biology, Chemistry, and Engineering}}}}\ (\bibinfo  {publisher} {Persues Books Publishing, L.L.C.},\ \bibinfo {address} {New York, USA},\ \bibinfo {year} {1994})\BibitemShut {NoStop}%
\bibitem [{\citenamefont {Bennett}\ \emph {et~al.}(1996)\citenamefont {Bennett}, \citenamefont {Brassard}, \citenamefont {Popescu}, \citenamefont {Schumacher}, \citenamefont {Smolin},\ and\ \citenamefont {Wooters}}]{BennettBrassardPopescuSchumacherSmolinWooters1996}%
  \BibitemOpen
  \bibfield  {author} {\bibinfo {author} {\bibfnamefont {Charles~H.}\ \bibnamefont {Bennett}}, \bibinfo {author} {\bibfnamefont {Gilles}\ \bibnamefont {Brassard}}, \bibinfo {author} {\bibfnamefont {Sandu}\ \bibnamefont {Popescu}}, \bibinfo {author} {\bibfnamefont {Benjamin}\ \bibnamefont {Schumacher}}, \bibinfo {author} {\bibfnamefont {John~A.}\ \bibnamefont {Smolin}}, \ and\ \bibinfo {author} {\bibfnamefont {William~K.}\ \bibnamefont {Wooters}},\ }\emph {\enquote {\bibinfo {title} {{Purification of Noisy Entanglement and Faithful Teleportation via Noisy Channels}},}\ }\href {\doibase 10.1103/physrevlett.76.722} {\bibfield  {journal} {\bibinfo  {journal} {Phys. Rev. Lett.}\ }\textbf {\bibinfo {volume} {76}},\ \bibinfo {pages} {722} (\bibinfo {year} {1996})},\ \Eprint {http://arxiv.org/abs/quant-ph/9511027} {arXiv:quant-ph/9511027}\BibitemShut {NoStop}%
\bibitem [{\citenamefont {Murao}\ \emph {et~al.}(1998)\citenamefont {Murao}, \citenamefont {Plenio}, \citenamefont {Popescu}, \citenamefont {Vedral},\ and\ \citenamefont {Knight}}]{MuraoPlenioPopescuVedralKnight1998}%
  \BibitemOpen
  \bibfield  {author} {\bibinfo {author} {\bibfnamefont {M.}~\bibnamefont {Murao}}, \bibinfo {author} {\bibfnamefont {M.~B.}\ \bibnamefont {Plenio}}, \bibinfo {author} {\bibfnamefont {S.}~\bibnamefont {Popescu}}, \bibinfo {author} {\bibfnamefont {V.}~\bibnamefont {Vedral}}, \ and\ \bibinfo {author} {\bibfnamefont {P.~L.}\ \bibnamefont {Knight}},\ }\emph {\enquote {\bibinfo {title} {Multiparticle entanglement purification protocols},}\ }\href {\doibase 10.1103/PhysRevA.57.R4075} {\bibfield  {journal} {\bibinfo  {journal} {Phys. Rev. A}\ }\textbf {\bibinfo {volume} {57}},\ \bibinfo {pages} {R4075} (\bibinfo {year} {1998})},\ \Eprint {http://arxiv.org/abs/quant-ph/9712045} {arXiv:quant-ph/9712045}\BibitemShut {NoStop}%
\bibitem [{\citenamefont {D\"ur}\ \emph {et~al.}(2003)\citenamefont {D\"ur}, \citenamefont {Aschauer},\ and\ \citenamefont {Briegel}}]{DuerAschauerBriegel2003}%
  \BibitemOpen
  \bibfield  {author} {\bibinfo {author} {\bibfnamefont {W.}~\bibnamefont {D\"ur}}, \bibinfo {author} {\bibfnamefont {H.}~\bibnamefont {Aschauer}}, \ and\ \bibinfo {author} {\bibfnamefont {H.-J.}\ \bibnamefont {Briegel}},\ }\emph {\enquote {\bibinfo {title} {Multiparticle entanglement purification for graph states},}\ }\href {\doibase 10.1103/PhysRevLett.91.107903} {\bibfield  {journal} {\bibinfo  {journal} {Phys. Rev. Lett.}\ }\textbf {\bibinfo {volume} {91}},\ \bibinfo {pages} {107903} (\bibinfo {year} {2003})},\ \Eprint {http://arxiv.org/abs/quant-ph/0303087} {arXiv:quant-ph/0303087}\BibitemShut {NoStop}%
\bibitem [{\citenamefont {D{\"u}r}\ and\ \citenamefont {Briegel}(2007)}]{DuerBriegel2007}%
  \BibitemOpen
  \bibfield  {author} {\bibinfo {author} {\bibfnamefont {Wolfgang}\ \bibnamefont {D{\"u}r}}\ and\ \bibinfo {author} {\bibfnamefont {Hans~J.}\ \bibnamefont {Briegel}},\ }\emph {\enquote {\bibinfo {title} {Entanglement purification and quantum error correction},}\ }\href {https://doi.org/10.1088/0034-4885/70/8/R03} {\bibfield  {journal} {\bibinfo  {journal} {Rep. Prog. Phys.}\ }\textbf {\bibinfo {volume} {70}},\ \bibinfo {pages} {1381} (\bibinfo {year} {2007})},\ \Eprint {http://arxiv.org/abs/0705.4165} {arXiv:0705.4165}\BibitemShut {NoStop}%
\bibitem [{\citenamefont {Horodecki}\ \emph {et~al.}(2022)\citenamefont {Horodecki}, \citenamefont {Rudnicki},\ and\ \citenamefont {{\.Z}yczkowski}}]{HorodeckiRudnickiZyczkowski2022}%
  \BibitemOpen
  \bibfield  {author} {\bibinfo {author} {\bibfnamefont {Pawe{\l}}\ \bibnamefont {Horodecki}}, \bibinfo {author} {\bibfnamefont {{\L}ukasz}\ \bibnamefont {Rudnicki}}, \ and\ \bibinfo {author} {\bibfnamefont {Karol}\ \bibnamefont {{\.Z}yczkowski}},\ }\emph {\enquote {\bibinfo {title} {Five open problems in quantum information theory},}\ }\href {https://doi.org/10.1103/PRXQuantum.3.01010} {\bibfield  {journal} {\bibinfo  {journal} {PRX Quantum}\ }\textbf {\bibinfo {volume} {3}},\ \bibinfo {pages} {010101} (\bibinfo {year} {2022})},\ \Eprint {http://arxiv.org/abs/2002.03233} {arXiv:2002.03233}\BibitemShut {NoStop}%
\bibitem [{\citenamefont {Huber}(2026)}]{huberdiscussion}%
  \BibitemOpen
  \bibfield  {author} {\bibinfo {author} {\bibfnamefont {Marcus}\ \bibnamefont {Huber}},\ }\href@noop {} {\emph {\enquote {\bibinfo {title} {Private communication},}\ }} (\bibinfo {year} {2026})\BibitemShut {NoStop}%
\bibitem [{\citenamefont {Rozgonyi}\ \emph {et~al.}(2025)\citenamefont {Rozgonyi}, \citenamefont {Sz\'echenyi}, \citenamefont {K\'alm\'an},\ and\ \citenamefont {Tam\'as}}]{RozgonyiSzechenyiKalmanKiss2025}%
  \BibitemOpen
  \bibfield  {author} {\bibinfo {author} {\bibfnamefont {\'Aron}\ \bibnamefont {Rozgonyi}}, \bibinfo {author} {\bibfnamefont {G\'abor}\ \bibnamefont {Sz\'echenyi}}, \bibinfo {author} {\bibfnamefont {Orsolya}\ \bibnamefont {K\'alm\'an}}, \ and\ \bibinfo {author} {\bibfnamefont {Kiss}\ \bibnamefont {Tam\'as}},\ }\href@noop {} {\emph {\enquote {\bibinfo {title} {{Practical scheme for efficient distillation of GHZ states}},}\ }}\Eprint {http://arxiv.org/abs/2501.12268} {arXiv:2501.12268} [quant-ph] (\bibinfo {year} {2025})\BibitemShut {NoStop}%
\end{thebibliography}%

\end{document}